\documentclass[letterpaper, 10 pt, conference]{ieeeconf} 

\IEEEoverridecommandlockouts                              
\usepackage{graphicx} 
\usepackage{amsmath} 
\usepackage{amssymb, bbm}  
\usepackage{lipsum}
\usepackage{multicol, multirow}
\usepackage{xcolor}
\usepackage{subfiles}
\usepackage[normalem]{ulem}
\usepackage{makecell}

\newcommand{\nuc}{\newcommand}
\nuc{\renuc}{\renewcommand}
\nuc{\dmo}{\DeclareMathOperator}

\nuc{\tcr}\textcopyright

\nuc{\noinsec}{\renewcommand{\theequation}{\arabic{section}.\arabic{equation}}}
\nuc{\noinapx}{\renewcommand{\theequation}{\Alph{section}.\arabic{equation}}}
\nuc{\noleminapx}{}
\nuc{\mysec}[1]{\section{#1}\setcounter{equation}{0}}

\nuc{\myapp}[1]{\section{#1}\setcounter{equation}{0}\noinapx\setcounter{lemma}{0}}

\newtheorem{alem}{Lemma}
\newtheorem{blem}{Lemma}
\newtheorem{clem}{Lemma}
\newtheorem{ass}{}

\nuc{\subsec}{\subsection}

\nuc{\ISd}{Itakura-Saito divergence}
\nuc{\pp}{point process}
\nuc{\bb}{Bhattacharyya bound}
\nuc{\cvd}{covariance density}
\nuc{\psd}{power spectral density}
\nuc{\BD}{Bregman divergence}
\nuc{\ifunc}{intensity function}
\nuc{\la}{local alternative}

\nuc{\Pp}{Point process}
\nuc{\Psp}{Poisson process}
\nuc{\rp}{renewal process}
\nuc{\Rnp}{Renewal process}
\nuc{\Hp}{Hawkes process}
\nuc{\HL}{Hawkes-Laguerre}
\nuc{\Lt}{Laplace transform}
\nuc{\iLt}{inverse Laplace transform}
\nuc{\iets}{interevent times}
\nuc{\Gd}{Gamma distributed}
\nuc{\lhf}{likelihood function}
\nuc{\lh}{likelihood}
\nuc{\tiv}{time-invariant}
\nuc{\tv}{time-varying}
\nuc{\Bhatt}{Bhattacharyya}
\nuc{\Bnl}{Bernoulli}
\nuc{\cpi}{counting process increment}

\nuc{\cme}{concentrated matrix exponential}
\nuc{\se}{self-exciting}
\nuc{\AWf}{Abate-Whitt framework}
\nuc{\mgf}{moment generating function}
\nuc{\KT}{Karamata Tauberian}
\nuc{\lyap}{Lyapunov }
\nuc{\iden}{identification }
\nuc{\sfr}{S$^4$}
\nuc{\sfv}{S$^5$}

\nuc{\rsi}{Riemann-Stieltjes integral}
\nuc{\RSi}{Riemann-Stieltjes integral}
\nuc{\RSc}{Riemann-Stieltjes convolution}
\nuc{\CSi}{Cauchy-Schwarz inequality}

\nuc{\ul}{\underline}

\nuc{\ra}{\rightarrow}
\nuc{\Ra}{\Rightarrow}
\nuc{\RA}{\Longrightarrow}
\nuc{\La}{\Leftarrow}
\nuc{\xra}[1]{\xrightarrow{#1}}
\nuc{\LRa}{\Leftrightarrow}
\nuc{\toi}{\to\infty}
\nuc{\trid}{\triangledown}
\nuc{\triu}{\triangleup}
\nuc{\bs}{\backslash}
\nuc{\convas}{\xra{a.s.}}
\nuc{\convp}{\xra{p}}
\nuc{\convd}{\xra{d}}
\nuc{\trieq}{\triangleq}

\nuc{\twn}{t_1^n}
\nuc{\xwn}{x_1^n}
\nuc{\Snoi}{\sum_{n=0}^\infty}
\nuc{\Snwi}{\sum_{n=1}^\infty}
\nuc{\swn}{\ssum1^n}
\nuc{\swN}{\ssum1^N}
\nuc{\swP}{\ssum1^P}
\nuc{\soN}{\ssum0^N}
\nuc{\swm}{\ssum1^m}
\nuc{\swM}{\ssum1^M}
\nuc{\siwn}{\ssum{i=1}^n}
\nuc{\siwnl}{\ssum{i=1}^{n_l}}
\nuc{\srwn}{\ssum{r=1}^n}
\nuc{\srwnl}{\ssum{r=1}^{n_l}}
\nuc{\srwnm}{\ssum{r=1}^{n_m}}
\nuc{\sawM}{\ssum{a=1}^M}
\nuc{\sbwM}{\ssum{b=1}^M}
\nuc{\siwm}{\ssum{i=1}^m}
\nuc{\skwm}{\ssum{k=1}^m}
\nuc{\siwM}{\ssum{i=1}^M}
\nuc{\skwM}{\ssum{k=1}^M}
\nuc{\siwK}{\ssum{i=1}^K}
\nuc{\sjwm}{\ssum{j=1}^m}
\nuc{\sjwM}{\ssum{j=1}^M}
\nuc{\slwM}{\ssum{l=1}^M}
\nuc{\smwM}{\ssum{m=1}^M}
\nuc{\sjwp}{\ssum{j=1}^p}
\nuc{\sjwP}{\ssum{j=1}^P}
\nuc{\swk}{\ssum1^k}
\nuc{\soi}{\ssum0^\infty}
\nuc{\swi}{\ssum1^\infty}
\nuc{\pwn}{\Pi_1^n}
\nuc{\pwk}{\Pi_1^k}
\nuc{\pwm}{\Pi_1^m}
\nuc{\prwn}{\Pi_{r=1}^n}
\nuc{\TwN}{T_1^N}

\nuc{\Stiltr}{\sum_{\ti<\tr}}
\nuc{\Sticlltrcm}{\sum_{\ticl<\trcm}}

\nuc{\Swn}{\sum_1^n}
\nuc{\Swm}{\sum_1^m}
\nuc{\SwM}{\sum_1^M}
\nuc{\Swk}{\sum_1^k}
\nuc{\Pwn}{\prod_1^n}
\nuc{\Pwm}{\prod_1^m}
\nuc{\Swp}{\sum_1^p}
\nuc{\SwP}{\sum_1^P}

\nuc{\intpi}{\int_{-\pi}^{\pi}}
\nuc{\intii}{\int_{-\infty}^{\infty}}
\nuc{\intoT}{\int_0^T}
\nuc{\intox}{\int_0^x}
\nuc{\intxi}{\int_x^\infty}
\nuc{\intti}{\int_t^\infty}
\nuc{\intit}{\int_{-\infty}^t}
\nuc{\intio}{\int_{-\infty}^0}
\nuc{\intiT}{\int_{-\infty}^T}
\nuc{\intot}{\int_0^t}
\nuc{\intou}{\int_0^u}
\nuc{\intoinfty}{\int_0^\infty}
\nuc{\intoi}{\int_0^\infty}
\nuc{\intmp}{\intinfty}

\nuc{\limto}{\lim_{t\to0}}
\nuc{\limTo}{\lim_{T\to0}}
\nuc{\limxo}{\lim_{x\to0}}
\nuc{\limso}{\lim_{s\to0}}
\nuc{\stoo}{s\to0}
\nuc{\stoi}{s\to\infty}
\nuc{\xtoi}{x\to\infty}
\nuc{\xtoo}{x\to0}
\nuc{\Ttoi}{T\to\infty}
\nuc{\ntoi}{n\to\infty}

\nuc{\limti}{\lim_{t\to\infty}}
\nuc{\limTi}{\lim_{T\to\infty}}
\nuc{\limxi}{\lim_{x\to\infty}}
\nuc{\limsi}{\lim_{s\to\infty}}

\nuc{\st}{\mbox{s.t. }}

\nuc{\hlamt}{\hat{\lambda}(t)}

\nuc{\nid}{n_i^\delta}
\nuc{\Nid}{N_i^\delta}

\nuc{\intsum}{\ssum0^\infty\int_{R_n(T)}}
\nuc{\ints}{\ssum0^\infty\int}

\nuc{\lt}{\left}
\nuc{\rt}{\right}

\nuc{\wos}{\frac1s}

\nuc{\wot}{\frac1t}
\nuc{\woT}{\frac1T}
\nuc{\wo}[1]{\frac1{#1}}
\nuc{\wox}{\wo}
\nuc{\haf}{{\frac12}}
\nuc{\mhaf}{{-\frac12}}

\nuc{\cH}{\mathcal{H}}
\nuc{\cB}{\mathcal{B}}
\nuc{\cD}{\mathcal{D}}
\nuc{\cN}{\mathcal{N}}
\nuc{\cP}{\mathcal{P}}
\nuc{\calj}{{\cal J}}
\nuc{\cHT}{\cH_T}
\nuc{\cO}{\mathcal{O}}
\nuc{\cL}{\mathcal{L}}
\nuc{\bR}{\mathbb R}
\nuc{\bC}{\mathbb C}
\nuc{\cl}{\mathcal{l}}
\nuc{\cI}{\mathcal{I}}
\nuc{\bbR}{\mathbb{R}}
\nuc{\bN}{\mathbb{N}}
\nuc{\bZ}{\mathbb{Z}}
\nuc{\bbw}[1]{\mathbbm{1}_{#1}}
\nuc{\N}{\mathfrak{N}}
\nuc{\n}{\mathfrak{n}}
\nuc{\bx}{{\bf x}}
\nuc{\by}{{\bf y}}
\nuc{\subs}{\subset}
\nuc{\wnb}{{\bf 1}}

\nuc{\eq}[1]{\begin{align*}#1\end{align*}}
\nuc{\eqn}[1]{\begin{align}#1\end{align}}
\nuc{\bmat}[1]{\begin{bmatrix}#1\end{bmatrix}}
\nuc{\iary}[2]{\begin{IEEEeqnarray*}{#1}#2\end{IEEEeqnarray*}}
\nuc{\iaryn}[2]{\begin{IEEEeqnarray}{#1}#2\end{IEEEeqnarray}}
\nuc{\subeq}[1]{\begin{subequations}#1\end{subequations}}
\nuc{\smat}[1]{\begin{smallmatrix}#1\end{smallmatrix}}
\nuc{\mat}[1]{\begin{matrix}#1\end{matrix}}
\nuc{\sqmat}[1]{\sqbra{\begin{matrix}#1\end{matrix}}}
\nuc{\chuz}[2]{\bra{\mat{#1\\#2}}}

\newcommand{\BNu}{\begin{enumerate}}
\newcommand{\ENu}{\end{enumerate}}
\newcommand{\Bit}{\begin{itemize}}
\newcommand{\Eit}{\end{itemize}}
\newcommand{\Bres}[1]{\begin{result}\label{#1}}
\newcommand{\Eres}{\end{result}}
\newcommand{\Bdef}[1]{\begin{defn}\label{#1}}
\newcommand{\Edef}{\end{defn}}
\newcommand{\Blem}[1]{\begin{lem}\label{#1}}
\newcommand{\Elem}{\end{lem}}
\newcommand{\Balem}[1]{\begin{alem}\label{#1}}
\newcommand{\Ealem}{\end{alem}}
\newcommand{\Bblem}[1]{\begin{blem}\label{#1}}
\newcommand{\Eblem}{\end{blem}}
\newcommand{\Bclem}[1]{\begin{clem}\label{#1}}
\newcommand{\Eclem}{\end{clem}}
\newcommand{\Brem}[1]{\begin{rem}\label{#1}}
\newcommand{\Erem}{\end{rem}}
\newcommand{\Bthm}[1]{\begin{thm}\label{#1}}
\newcommand{\Ethm}{\end{thm}}
\newcommand{\Bass}[1]{\begin{ass}\label{#1}}
\newcommand{\Eass}{\end{ass}}
\newcommand{\Balg}[1]{\begin{alg}\label{#1}}
\newcommand{\Ealg}{\end{alg}}
\newcommand{\Bprop}[1]{\begin{prop}\label{#1}}
\newcommand{\Eprop}{\end{prop}}
\newcommand{\Bcor}[1]{\begin{coro}\label{#1}}
\newcommand{\Ecor}{\end{coro}}
\newcommand{\Bpf}{\begin{pf*}}
\newcommand{\Epf}{{\hfill$\square$}\end{pf*}}
\newcommand{\Bsmat}[1]{\begin{smallmatrix}#1}
\newcommand{\Esmat}{\end{smallmatrix}}
\newcommand{\Bmat}[1]{\begin{matrix}#1}
\newcommand{\Emat}{\end{matrix}}
\nuc{\Brmk}{\begin{remark}}
\nuc{\Brmks}{\begin{remarks}}
\nuc{\Ermk}{\end{remark}}
\nuc{\Ermks}{\end{remarks}}

\nuc{\emu}[1]{e^{-#1}}

\nuc{\theo}[1]{\begin{thm}#1\end{thm}}
\nuc{\defi}[1]{\begin{defn}#1\end{defn}}
\nuc{\pro}[1]{\begin{proof}#1\end{proof}}
\nuc{\cas}[1]{\begin{cases}#1\end{cases}}
\nuc{\arr}[2]{\begin{array}{#1}#2\end{array}}
\nuc{\bra}[1]{\left(#1\right)}
\nuc{\sqbra}[1]{\left[#1\right]}
\nuc{\ang}[1]{\langle#1\rangle}
\nuc{\floor}[1]{\lfloor#1\rfloor}
\nuc{\Ver}[1]{\lVert#1\rVert}
\nuc{\ver}[1]{\lvert#1\rvert}
\nuc{\Bver}[1]{\Bigl\vert#1\Bigr\vert}
\nuc{\bver}[1]{\bigl\vert#1\bigr\vert}
\nuc{\BVer}[1]{\Bigl\Vert#1\Bigr\Vert}
\nuc{\bVer}[1]{\bigl\Vert#1\bigr\Vert}
\nuc{\ssum}[1]{{\textstyle\sum}_{#1}}
\nuc{\sprod}[1]{{\textstyle\prod}_{#1}}
\nuc{\inv}[1]{#1^{-1}}
\nuc{\ddt}[1]{\frac{d#1}{dt}}
\nuc{\ddx}[2]{\frac{d#1}{d#2}}
\nuc{\ppt}[1]{\frac{\partial#1}{\partial t}}
\nuc{\ppx}[2]{\frac{\partial#1}{\partial#2}}

\nuc{\sbij}[1]{#1_{ij}}
\nuc{\sbbij}[1]{\bar #1_{ij}}
\nuc{\sbi}[1]{#1_{i}}
\nuc{\sbbi}[1]{\bar #1_{i}}
\nuc{\sbj}[1]{#1_{j}}
\nuc{\sbbj}[1]{\bar #1_{j}}
\nuc{\nfd}[1]{#1^{(n)}}
\nuc{\limo}[1]{\lim_{#1\to0}}
\nuc{\limi}[1]{\lim_{#1\to\infty}}

\nuc{\grad}{\bigtriangledown}
\nuc{\gradx}[1]{\bigtriangledown_{#1}}
\nuc{\hessx}[1]{\bigtriangledown^2_{#1}}

\nuc{\idc}[1]{\mathbbm{1}_{#1}}

\newcommand{\blist}[1]{\begin{itemize}#1\end{itemize}}
\nuc{\itum}[1]{\item[(#1)]}
\nuc{\ita}{\itum{a}}
\nuc{\itb}{\itum{b}}
\nuc{\itc}{\itum{c}}
\nuc{\itd}{\itum{d}}
\nuc{\ite}{\itum{e}}
\nuc{\itf}{\itum{f}}
\nuc{\itg}{\itum{g}}
\nuc{\ith}{\itum{h}}

\DeclareMathOperator{\E}{E}

\DeclareMathOperator{\trace}{tr}

\DeclareMathOperator*{\argmin}{arg\,min}

\nuc{\ai}{a_i}
\nuc{\ak}{a_k}
\nuc{\aj}{a_j}
\nuc{\aij}{a_{ij}}
\nuc{\akl}{a_{kl}}
\nuc{\akj}{a_{kj}}
\nuc{\abi}{\bar a_i}
\nuc{\abk}{\bar a_k}
\nuc{\abj}{\bar a_j}
\nuc{\abij}{\bar a_{ij}}
\nuc{\abkj}{\bar a_{kj}}
\nuc{\alpm}{\alpha_{m}}
\nuc{\alpi}{\alpha_{i}}
\nuc{\alpk}{\alpha_{k}}
\nuc{\alpj}{\alpha_{j}}
\nuc{\alpij}{\alpha_{ij}}
\nuc{\alpkj}{\alpha_{kj}}
\nuc{\alpab}{\alpha_{ab}}
\nuc{\Ahw}{\hat A_1}
\nuc{\Aho}{\hat A_0}
\nuc{\ajcl}{a_{j,l}}
\nuc{\ajclcm}{a_{j,l,m}}
\nuc{\Ahat}{\hat A}
\nuc{\Ab}{A_b}

\nuc{\Ah}{\hat{A}}
\nuc{\alp}{\alpha}

\nuc{\Bi}{B_i}
\nuc{\Bj}{B_j}
\nuc{\Bij}{B_{ij}}
\nuc{\Bkj}{B_{kj}}
\nuc{\bij}{b_{ij}}
\nuc{\bkj}{b_{kj}}
\nuc{\Bbi}{\bar B_i}
\nuc{\Bbj}{\bar B_j}
\nuc{\Bbij}{\bar B_{ij}}
\nuc{\Bbkj}{\bar B_{kj}}
\nuc{\beti}{\beta_i}
\nuc{\betj}{\beta_j}
\nuc{\betk}{\beta_k}
\nuc{\betaa}{\beta_a}
\nuc{\betb}{\beta_b}
\nuc{\betij}{\beta_{ij}}
\nuc{\betkl}{\beta_{kl}}
\nuc{\betkj}{\beta_{kj}}
\nuc{\betab}{\beta_{ab}}
\nuc{\betjcl}{\beta_{j,l}}
\nuc{\betjclcm}{\beta_{j,l,m}}
\nuc{\Bhw}{\hat B_1}
\nuc{\Bho}{\hat B_0}

\nuc{\ci}{c_i}
\nuc{\cj}{c_j}
\nuc{\ck}{c_k}
\nuc{\ca}{c_a}
\nuc{\cb}{c_b}

\nuc{\cm}{c_m}
\nuc{\cij}{c_{ij}}
\nuc{\ckj}{c_{kj}}
\nuc{\cab}{c_{ab}}
\nuc{\cbi}{\bar c_i}
\nuc{\cbj}{\bar c_j}
\nuc{\Ci}{C_i}
\nuc{\Cj}{C_j}
\nuc{\Cij}{C_{ij}}
\nuc{\Cbi}{\bar C_i}
\nuc{\Cbj}{\bar C_j}
\nuc{\Chw}{\hat C_1}
\nuc{\Cho}{\hat C_0}
\nuc{\cwt}{c_{12}}
\nuc{\Chat}{\hat C}

\nuc{\Di}{D_i}
\nuc{\Dj}{D_j}
\nuc{\Dbi}{\bar D_i}
\nuc{\Dbj}{\bar D_j}
\nuc{\Dij}{D_{ij}}
\nuc{\Dkj}{D_{kj}}
\nuc{\Dab}{D_{ab}}
\nuc{\dij}{d_{ij}}
\nuc{\dab}{d_{ab}}
\nuc{\dkj}{d_{kj}}
\nuc{\Dbij}{\bar D_{ij}}
\nuc{\Dbkj}{\bar D_{kj}}
\nuc{\Dbab}{\bar D_{ab}}
\nuc{\DT}{D_T}

\nuc{\et}{e_t}
\nuc{\ep}{\epsilon}

\nuc{\fk}{f_k}
\nuc{\Fi}{F_i}
\nuc{\Fk}{F_k}
\nuc{\Fj}{F_j}
\nuc{\Fa}{F_a}
\nuc{\Fb}{F_b}
\nuc{\Fm}{F_m}
\nuc{\Ft}{F_t}
\nuc{\Fbi}{\bar F_i}
\nuc{\Fbj}{\bar F_j}
\nuc{\Fba}{\bar F_a}
\nuc{\Fbb}{\bar F_b}
\nuc{\Fij}{F_{ij}}
\nuc{\Fkj}{F_{kj}}
\nuc{\Fab}{F_{ab}}
\nuc{\fij}{f_{ij}}
\nuc{\fkj}{f_{kj}}
\nuc{\fab}{f_{ab}}
\nuc{\Fbij}{\bar F_{ij}}
\nuc{\Fbkj}{\bar F_{kj}}
\nuc{\Fbab}{\bar F_{ab}}
\nuc{\Fwtu}{F_{12}}
\nuc{\Fbwtu}{\bar F_{12}}
\nuc{\Fpct}{F_{p,t}}
\nuc{\Fh}{\hat F}
\nuc{\Fhat}{\hat F}

\nuc{\gam}{\gamma}
\nuc{\Gam}{\Gamma}
\nuc{\gami}{\gam_i}
\nuc{\gamk}{\gam_k}
\nuc{\gamj}{\gam_j}
\nuc{\gamij}{\gam_{ij}}
\nuc{\gamkj}{\gam_{kj}}
\nuc{\gamab}{\gam_{ab}}
\nuc{\Gi}{G_i}
\nuc{\Gk}{G_k}
\nuc{\Gj}{G_j}
\nuc{\Ga}{G_a}
\nuc{\Gb}{G_b}
\nuc{\Gbi}{\bar G_i}
\nuc{\Gbj}{\bar G_j}
\nuc{\Gij}{G_{ij}}
\nuc{\Gkj}{G_{kj}}
\nuc{\Gab}{G_{ab}}
\nuc{\Gbab}{\bar G_{ab}}
\nuc{\Gbij}{\bar G_{ij}}
\nuc{\Gbkj}{\bar G_{kj}}
\nuc{\Gwtu}{G_{12}}
\nuc{\Gbwtu}{\bar G_{12}}
\nuc{\bGwt}{\bar G_{12}}
\nuc{\Gwt}{G_{12}}

\nuc{\bgwt}{\bar g_{12}}

\nuc{\gwt}{g_{12}}

\nuc{\hi}{h_i}
\nuc{\hk}{h_k}
\nuc{\hj}{h_j}
\nuc{\ha}{h_a}
\nuc{\hb}{h_b}
\nuc{\hd}{h_d}
\nuc{\hbi}{\bar h_i}
\nuc{\hbk}{\bar h_k}
\nuc{\hbj}{\bar h_j}
\nuc{\hba}{\bar h_a}
\nuc{\hbb}{\bar h_b}
\nuc{\hij}{h_{ij}}
\nuc{\hbij}{\bar h_{ij}}
\nuc{\hkj}{h_{kj}}
\nuc{\hbkj}{\bar h_{kj}}
\nuc{\hab}{h_{ab}}
\nuc{\hbab}{\bar h_{ab}}
\nuc{\hjcm}{h_{j,m}}
\nuc{\hjcl}{h_{j,l}}
\nuc{\hjclcm}{h_{j,l,m}}
\nuc{\Hi}{H_i}
\nuc{\Hk}{H_k}
\nuc{\Hj}{H_j}
\nuc{\Hbi}{\bar H_i}
\nuc{\Hbk}{\bar H_k}
\nuc{\Hbj}{\bar H_j}
\nuc{\Hij}{H_{ij}}
\nuc{\Hbij}{\bar H_{ij}}
\nuc{\Hkj}{H_{kj}}
\nuc{\Hbkj}{\bar H_{kj}}
\nuc{\Hab}{H_{kj}}
\nuc{\Hbab}{\bar H_{ab}}
\nuc{\Hwtu}{H_{12}}
\nuc{\Hbwtu}{\bar H_{12}}
\nuc{\Hinf}{\cH_\infty}
\nuc{\cHTicj}{\cH_{T,i,j}}
\nuc{\calhr}{{\cal H}_R}

\nuc{\jw}{\jmath\omega}

\nuc{\kapij}{\kappa_{ij}}
\nuc{\kapbij}{\bar\kappa_{ij}}
\nuc{\kapkj}{\kappa_{kj}}
\nuc{\kapbkj}{\bar\kappa_{kj}}
\nuc{\kapab}{\kappa_{ab}}
\nuc{\kapbab}{\bar\kappa_{ab}}
\nuc{\kapji}{\kappa_{ji}}
\nuc{\kapbji}{\bar\kappa_{ji}}
\nuc{\Ki}{K_i}
\nuc{\Kk}{K_k}
\nuc{\Kj}{K_j}
\nuc{\Kt}{K_t}
\nuc{\Kbi}{\bar K_i}
\nuc{\Kbk}{\bar K_k}
\nuc{\Kbj}{\bar K_j}
\nuc{\Kij}{K_{ij}}
\nuc{\Kbij}{\bar K_{ij}}
\nuc{\Kkj}{K_{kj}}
\nuc{\Kbkj}{\bar K_{kj}}
\nuc{\Kab}{K_{ab}}
\nuc{\Kbab}{\bar K_{ab}}
\nuc{\Kji}{K_{ji}}
\nuc{\Kbji}{\bar K_{ji}}
\nuc{\Kwtu}{K_{12}}
\nuc{\Kbwtu}{\bar K_{12}}
\nuc{\calhk}{{\cal H}_K}

\nuc{\lam}{\lambda}
\nuc{\Lam}{\Lambda}
\nuc{\lami}{\lambda_{i}}
\nuc{\lamk}{\lambda_{k}}
\nuc{\lamj}{\lambda_{j}}
\nuc{\lama}{\lambda_{a}}
\nuc{\lamb}{\lambda_{b}}
\nuc{\lamm}{\lambda_{m}}
\nuc{\laml}{\lambda_{l}}
\nuc{\lamij}{\lambda_{ij}}
\nuc{\lamkj}{\lambda_{kj}}
\nuc{\lamab}{\lambda_{ab}}
\nuc{\lamicj}{\lambda_{i,j}}
\nuc{\lambij}{\bar\lambda_{ij}}
\nuc{\lambab}{\bar\lambda_{ab}}
\nuc{\Lami}{\Lambda_{i}}
\nuc{\Lamk}{\Lambda_{k}}
\nuc{\Lama}{\Lambda_{a}}
\nuc{\Lamj}{\Lambda_{j}}
\nuc{\Lamij}{\Lambda_{ij}}
\nuc{\Lamkj}{\Lambda_{kj}}
\nuc{\Lamab}{\Lambda_{ab}}
\nuc{\Lamicj}{\Lambda_{i,j}}
\nuc{\Lambij}{\bar\Lambda_{ij}}

\nuc{\lamTt}{\lam_t^T}
\nuc{\lamtgT}{\lam_{t|T}}
\nuc{\lamTu}{\lam_u^T}
\nuc{\lamugT}{\lam_{u|T}}
\nuc{\lamicjclcm}{\lam_{i,j,l,m}}
\nuc{\lamTtcu}{\lam_{t,u}^T}
\nuc{\lamtcugT}{\lam_{t,u|T}}
\nuc{\lamTtpw}{\lam^T_{t+1}}
\nuc{\lamtpwgT}{\lam_{t+1|T}}
\nuc{\LamTt}{\Lam_t^T}
\nuc{\LamtgT}{\Lam_{t|T}}
\nuc{\LamTu}{\Lam_u^T}
\nuc{\LamugT}{\Lam_{u|T}}
\nuc{\LamTtpw}{\Lam^T_{t+1}}
\nuc{\LamtpwgT}{\Lam_{t+1|T}}
\nuc{\LamTtcu}{\Lam_{t,u}^T}
\nuc{\LamtcugT}{\Lam_{t,u|T}}

\nuc{\Li}{L_i}
\nuc{\Lk}{L_k}
\nuc{\Lj}{L_j}
\nuc{\cLa}{\cL_a}
\nuc{\cLb}{\cL_b}

\nuc{\Lb}{L_b}
\nuc{\Lbi}{\bar L_i}
\nuc{\Lbk}{\bar L_k}
\nuc{\Lbj}{\bar L_j}
\nuc{\Lij}{L_{ij}}
\nuc{\Lkj}{L_{kj}}
\nuc{\Lab}{L_{ab}}
\nuc{\Licj}{L_{i,j}}
\nuc{\Lbij}{\bar L_{ij}}
\nuc{\Lbkj}{\bar L_{kj}}
\nuc{\Lwtu}{L_{12}}
\nuc{\Lbwtu}{\bar L_{12}}
\nuc{\LTtn}{L(T,n;t_1^n)}
\nuc{\mi}{m_i}
\nuc{\mk}{m_k}
\nuc{\mj}{m_j}
\nuc{\mij}{m_{ij}}
\nuc{\mkj}{m_{kj}}
\nuc{\mab}{m_{ab}}
\nuc{\Mij}{M_{ij}}
\nuc{\Mkj}{M_{kj}}
\nuc{\Mab}{M_{ab}}
\nuc{\Mfij}{M_{\fij}}
\nuc{\mui}{\mu_i}
\nuc{\muk}{\mu_k}
\nuc{\muj}{\mu_j}
\nuc{\mua}{\mu_a}
\nuc{\mub}{\mu_b}
\nuc{\muij}{\mu_{ij}}
\nuc{\mukj}{\mu_{kj}}
\nuc{\muab}{\mu_{ab}}
\nuc{\mbi}{\bar m_i}
\nuc{\mbk}{\bar m_k}
\nuc{\mbj}{\bar m_j}
\nuc{\mbij}{\bar m_{ij}}
\nuc{\mbkj}{\bar m_{kj}}

\nuc{\NoT}{N_0^T}
\nuc{\nut}{\nu_t}
\nuc{\nl}{n_l}
\nuc{\nm}{n_m}
\nuc{\Nt}{N_t}
\nuc{\NT}{N_T}
\nuc{\nT}{n_T}
\nuc{\nicj}{n_{i,j}}
\nuc{\Nu}{N_u}
\nuc{\Nucm}{N_{u,m}}
\nuc{\Nucl}{N_{u,l}}
\nuc{\Ntcl}{N_{t,l}}
\nuc{\Not}{N_0^t}
\nuc{\Notm}{N_0^{t_-}}

\nuc{\ome}{\omega}
\nuc{\Ome}{\Omega}
\nuc{\omekl}{\ome_{kl}}
\nuc{\omeal}{\ome_{al}}
\nuc{\omet}{\ome_t}
\nuc{\Omeicj}{\Ome_{i,j}}
\nuc{\Omebicj}{\bar\Ome_{i,j}}

\nuc{\pbi}{\bar p_i}
\nuc{\pbk}{\bar p_k}
\nuc{\pbj}{\bar p_j}
\nuc{\pba}{\bar p_a}
\nuc{\pbb}{\bar p_b}
\nuc{\pik}{\pi_k}
\nuc{\pj}{p_j}
\nuc{\pk}{p_k}
\nuc{\pa}{p_a}
\nuc{\pb}{p_b}
\nuc{\pij}{p_{ij}}
\nuc{\pkj}{p_{kj}}
\nuc{\pab}{p_{ab}}
\nuc{\pbar}{\bar p}
\nuc{\pbij}{\bar p_{ij}}
\nuc{\pbkj}{\bar p_{kj}}
\nuc{\pbab}{\bar p_{ab}}
\nuc{\ptij}{\tilde p_{ij}}
\nuc{\pwtu}{p_{12}}
\nuc{\pbwtu}{\bar p_{12}}

\nuc{\Picj}{P_{i,j}}
\nuc{\PTicj}{P^T_{i,j}}
\nuc{\PTicipw}{P^T_{i,i+w}}
\nuc{\PTicipk}{P^T_{i,i+k}}

\nuc{\Pt}{P_t}
\nuc{\Pu}{P_u}
\nuc{\Ptcu}{P_{t,u}}
\nuc{\Ptck}{P_{t,k}}
\nuc{\PTtcu}{P^T_{t,u}}
\nuc{\PtcugT}{P_{t,u|T}}
\nuc{\Pstcu}{P^s_{t,u}}
\nuc{\Ptcugs}{P_{t,u|s}}
\nuc{\PTtctpw}{P^T_{t,t+w}}
\nuc{\PTtctpk}{P^T_{t,t+k}}
\nuc{\Ptt}{P^t_t}
\nuc{\Ptgt}{P_{t|t}}
\nuc{\PtgT}{P_{t|T}}
\nuc{\Pst}{P^s_t}
\nuc{\Ptgs}{P_{t|s}}
\nuc{\PTt}{P^T_t}
\nuc{\Ptmwt}{P^{t-1}_t}
\nuc{\Ptgtmw}{P_{t|t-1}}
\nuc{\Po}{P_0}
\nuc{\Poi}{P_0^{-1}}

\nuc{\phiad}{\phi_{ad}}
\nuc{\phiba}{\phi_{ba}}
\nuc{\phiaa}{\phi_{aa}}
\nuc{\phibd}{\phi_{bd}}
\nuc{\phicd}{\phi_{cd}}
\nuc{\phiadh}{\hat\phi_{ad}}
\nuc{\phibah}{\hat\phi_{ba}}
\nuc{\phiaah}{\hat\phi_{aa}}
\nuc{\phibdh}{\hat\phi_{bd}}
\nuc{\phicdh}{\hat\phi_{cd}}

\nuc{\bpwt}{\bar p_{12}}
\nuc{\pwt}{p_{12}}

\nuc{\Qhat}{\hat Q}
\nuc{\Qbi}{\bar Q_i}
\nuc{\Qbk}{\bar Q_k}
\nuc{\Qbj}{\bar Q_j}
\nuc{\Qj}{Q_j}
\nuc{\Qij}{Q_{ij}}
\nuc{\Qkj}{Q_{kj}}
\nuc{\Qab}{Q_{ab}}
\nuc{\Qm}{Q_m}
\nuc{\Qjclcm}{Q_{j,l,m}}
\nuc{\Qbij}{\bar Q_{ij}}
\nuc{\Qwtu}{Q_{12}}
\nuc{\Qbwtu}{\bar Q_{12}}
\nuc{\qbi}{\bar q_i}
\nuc{\qbk}{\bar q_k}
\nuc{\qbj}{\bar q_j}
\nuc{\qba}{\bar q_a}
\nuc{\qbb}{\bar q_b}
\nuc{\qj}{q_j}
\nuc{\qij}{q_{ij}}
\nuc{\qbij}{\bar q_{ij}}
\nuc{\qkj}{q_{kj}}
\nuc{\qbkj}{\bar q_{kj}}
\nuc{\qab}{q_{ab}}
\nuc{\qbab}{\bar q_{ab}}
\nuc{\qwtu}{q_{12}}
\nuc{\qbwtu}{\bar q_{12}}
\nuc{\phiT}{\phi_T}
\nuc{\bqwt}{\bar q_{12}}

\nuc{\qwt}{q_{12}}

\nuc{\Qv}{Q_v}
\nuc{\Qep}{Q_\ep}
\nuc{\Qw}{Q_w}

\nuc{\Rhat}{\hat R}
\nuc{\Rbi}{\bar R_i}
\nuc{\Rbj}{\bar R_j}
\nuc{\Ri}{R_i}
\nuc{\Rj}{R_j}
\nuc{\Rt}{R_t}
\nuc{\Rij}{R_{ij}}
\nuc{\Rkj}{R_{kj}}
\nuc{\Rab}{R_{ab}}
\nuc{\Rjclcm}{R_{j,l,m}}
\nuc{\rij}{r_{ij}}
\nuc{\Rbij}{\bar R_{ij}}
\nuc{\Rbkj}{\bar R_{kj}}
\nuc{\Rbab}{\bar R_{ab}}
\nuc{\Rwtu}{R_{12}}
\nuc{\Rbwtu}{\bar R_{12}}
\nuc{\rhoT}{\rho_T}
\nuc{\rhoi}{\rho_{i}}
\nuc{\rhok}{\rho_{k}}
\nuc{\rhoj}{\rho_{j}}
\nuc{\rhoa}{\rho_{a}}
\nuc{\rhob}{\rho_{b}}
\nuc{\rhoij}{\rho_{ij}}
\nuc{\rhokj}{\rho_{kj}}
\nuc{\rhoab}{\rho_{ab}}
\nuc{\Rect}{R_{e,t}}

\nuc{\Sig}{\Sigma}
\nuc{\sig}{\sigma}
\nuc{\Si}{S_i}
\nuc{\Sk}{S_k}
\nuc{\Sj}{S_j}
\nuc{\Sa}{S_a}
\nuc{\Sb}{S_b}
\nuc{\Sbar}{\bar S}
\nuc{\Sbi}{\bar S_i}
\nuc{\Sbk}{\bar S_k}
\nuc{\Sbj}{\bar S_j}
\nuc{\Sba}{\bar S_a}
\nuc{\Sbb}{\bar S_b}
\nuc{\Sij}{S_{ij}}
\nuc{\Sbij}{\bar S_{ij}}
\nuc{\Skj}{S_{kj}}
\nuc{\Sbkj}{\bar S_{kj}}
\nuc{\Sab}{S_{ab}}
\nuc{\Sbab}{\bar S_{ab}}
\nuc{\Swtu}{S_{12}}
\nuc{\Sww}{S_{11}}
\nuc{\Swo}{S_{10}}
\nuc{\Sow}{S_{01}}
\nuc{\Soo}{S_{00}}
\nuc{\Sbwtu}{\bar S_{12}}
\nuc{\Sigxx}{\ssum{xx}}
\nuc{\Sigx}{\ssum{x}}
\nuc{\Sigy}{\ssum{y}}
\nuc{\Sigxy}{\ssum{xy}}
\nuc{\Sigyx}{\ssum{yx}}
\nuc{\Sigyy}{\ssum{yy}}
\nuc{\Sxx}{S_{xx}}
\nuc{\Sx}{S_{x}}
\nuc{\Sy}{S_{x}}
\nuc{\Sxy}{S_{xy}}
\nuc{\Syx}{S_{yx}}
\nuc{\Syy}{S_{yy}}
\nuc{\Sjclcm}{S_{j,l,m}}
\nuc{\sigi}{\sigma_i}
\nuc{\sigk}{\sigma_k}
\nuc{\siga}{\sigma_a}
\nuc{\sigb}{\sigma_b}
\nuc{\Sw}{S_w}
\nuc{\Swf}{S_{w,f}}
\nuc{\Swb}{S_{w,b}}

\nuc{\Tb}{\bar T}
\nuc{\tn}{t_n}
\nuc{\tr}{t_r}
\nuc{\ti}{t_i}
\nuc{\ticl}{t_{i,l}}
\nuc{\trcm}{t_{r,m}}
\nuc{\twnm}{t_1^{n_m}}
\nuc{\twnl}{t_1^{n_l}}
\nuc{\twclnl}{t_{1,l}^{n_l}}
\nuc{\trmw}{t_{r-1}}
\nuc{\trpw}{t_{r+1}}
\nuc{\thehw}{\hat\theta_1}
\nuc{\theho}{\hat\theta_0}
\nuc{\thek}{\theta_k}
\nuc{\thej}{\theta_j}
\nuc{\thea}{\theta_a}
\nuc{\theb}{\theta_b}
\nuc{\theab}{\theta_{ab}}
\nuc{\thekj}{\theta_{kj}}
\nuc{\dtri}{\tr - \ti}
\nuc{\dtrcmicl}{\trcm - \ticl}
\nuc{\dTtr}{T - \tr}
\nuc{\dTti}{T - \ti}
\nuc{\dTticl}{T - \ticl}
\nuc{\dtr}{\trpw-\tr}
\nuc{\dtrcm}{t_{r+1,m}-\trcm}
\nuc{\tauwnicj}{\tau_1^{\nicj}}
\nuc{\tauicj}{\tau_{i,j}}
\nuc{\tauicjcl}{\tau_{i,j,l}}

\nuc{\ui}{u_i}
\nuc{\uk}{u_k}
\nuc{\uj}{u_j}
\nuc{\uij}{u_{ij}}
\nuc{\ukj}{u_{kj}}
\nuc{\uab}{u_{ab}}
\nuc{\ubi}{\bar u_i}
\nuc{\ubj}{\bar u_j}
\nuc{\Ui}{U_i}
\nuc{\Uj}{U_j}
\nuc{\Uij}{U_{ij}}
\nuc{\Ubi}{\bar U_i}
\nuc{\Ubj}{\bar U_j}

\nuc{\vi}{v_i}
\nuc{\vj}{v_j}
\nuc{\vij}{v_{ij}}
\nuc{\vbi}{\bar v_i}
\nuc{\vbj}{\bar v_j}
\nuc{\Vi}{V_i}
\nuc{\Vj}{V_j}
\nuc{\Vij}{V_{ij}}
\nuc{\Vbi}{\bar V_i}
\nuc{\Vbj}{\bar V_j}

\nuc{\wt}{w_t}
\nuc{\wtmw}{w_{t-1}}
\nuc{\wtpw}{w_{t+1}}
\nuc{\wicj}{w_{i,j}}
\nuc{\wicjclcm}{w_{i,j,l,m}}
\nuc{\wbt}{w_{b,t}}
\nuc{\wbtpw}{w_{b,t+1}}
\nuc{\wb}{w_b}

\nuc{\Xhat}{\hat X}

\nuc{\xt}{x_t}
\nuc{\xtpw}{x_{t+1}}
\nuc{\xtmw}{x_{t-1}}
\nuc{\xo}{x_0}
\nuc{\xw}{x_1}
\nuc{\xto}{x_2}
\nuc{\xT}{x_T}

\nuc{\xTt}{x^T_t}
\nuc{\xtt}{x^t_t}
\nuc{\xtmwt}{x^{t-1}_t}

\nuc{\xtil}{\tilde{x}}
\nuc{\xtilt}{\tilde{x}_t}
\nuc{\xtilu}{\tilde{x}_u}
\nuc{\xtiltgs}{\tilde{x}_{t|s}}
\nuc{\xtilugs}{\tilde{x}_{u|s}}
\nuc{\xtiltgt}{\tilde{x}_{t|t}}
\nuc{\xtiltgT}{\tilde{x}_{t|T}}
\nuc{\xtilugT}{\tilde{x}_{u|T}}
\nuc{\xtiltgtmw}{\tilde{x}_{t|t-1}}
\nuc{\xhat}{\hat{x}}
\nuc{\xhatt}{\hat{x}_t}
\nuc{\xhattgs}{\hat{x}_{t|s}}
\nuc{\xhattgtmw}{\hat{x}_{t|t-1}}
\nuc{\xhattt}{\hat{x}_t^t}
\nuc{\xhattgt}{\hat{x}_{t|t}}
\nuc{\xhattgT}{\hat{x}_{t|T}}
\nuc{\xhattmwgtmw}{\xhat_{t-1|t-1}}

\nuc{\xht}{\xh_t}
\nuc{\xh}{\hat{x}}
\nuc{\Xh}{\hat{X}}

\nuc{\yt}{y_t}
\nuc{\yr}{y_r}
\nuc{\ywn}{y_1^n}
\nuc{\ywT}{y_1^T}
\nuc{\yws}{y_1^s}
\nuc{\Yp}{Y_p}
\nuc{\Yf}{Y_f}
\nuc{\ytpw}{y_{t+1}}
\nuc{\ytmw}{y_{t-1}}
\nuc{\zt}{z_t}
\nuc{\ztpw}{z_{t+1}}
\nuc{\zit}{\zeta_t}
\nuc{\zitp}{\zeta_{t+1}}
\nuc{\clrb}[1]{{\color{blue}{#1}}}
\nuc{\clrr}[1]{{\color{red}{#1}}}

\nuc{\wpw}{\mbox{w.p.1.}}

\dmo{\T}{T}
\nuc{\bP}{\mathbb{P}}
\nuc{\cF}{\mathcal{F}}
\nuc{\cHtm}{\cH_{t-}}
\nuc{\cHsm}{\cH_{s-}}
\nuc{\cHom}{\cH_{0-}}
\renuc{\cI}{\mathbf{1}}

\renuc{\choose}[2]{\bra{\smat{#1\\#2}}}

\nuc{\intitm}{\int_{-\infty}^{t-}}

\nuc{\umck}{unit-mass causal kernel}

\nuc{\limdo}{\lim_{\delta\to0}}
\nuc{\limdoo}{\lim_{\delta\to0}\wo\delta}

\renuc{\thehw}{\hat\theta_T^{(1)}}
\nuc{\wnt}{\mathcal{I}_{[0,\infty)}(t)}

\nuc{\WkTo}{W_{k,T}^0}
\nuc{\Sigthek}{\Sig_{\thek}}
\nuc{\Sigalpk}{\Sig_{\alpk}}
\nuc{\Sigck}{\Sig_{\ck}}

\nuc{\chij}{\chi_j}
\nuc{\phib}{\bar\phi}
\nuc{\phibo}{\bar\phi^{(0)}}
\nuc{\Mt}{M_t}
\nuc{\lamtilh}{\tilde\lam^{(h)}}
\nuc{\lamtilw}{\tilde\lam^{(1)}}
\nuc{\lamtilo}{\tilde\lam^{(0)}}
\nuc{\cw}{c^{(1)}}
\nuc{\alpw}{\alpha^{(1)}}
\nuc{\co}{c^{(0)}}
\nuc{\ch}{c^{(h)}}
\nuc{\chw}{\hat c_T^{(1)}}
\nuc{\csw}{c_*^{(1)}}

\nuc{\chijph}{\chi_{j'}^{(h)}}
\nuc{\chitilh}{\tilde\chi^{(h)}}
\nuc{\chih}{\hat\chi}
\nuc{\chitiljh}{\tilde\chi_j^{(h)}}
\nuc{\chitiljph}{\tilde\chi_{j'}^{(h)}}
\nuc{\chitilw}{\tilde\chi_1}
\nuc{\chitilwT}{\tilde\chi_1^\top}
\nuc{\chitiljw}{\tilde\chi_j^{(1)}}
\nuc{\chitilo}{\tilde\chi^{(0)}}
\nuc{\chitiljo}{\tilde\chi_j^{(0)}}

\renuc{\alph}{\hat\alp}
\nuc{\alphw}{\hat\alpha_T^{(1)}}
\nuc{\alpsw}{\alpha_*^{(1)}}

\nuc{\alphwT}{\hat\alpha_T^{(1)\top}}

\nuc{\qjh}{q_j^{(h)}}
\nuc{\qjph}{q_{j'}^{(h)}}
\nuc{\qjw}{q_j^{(1)}}
\nuc{\qjo}{q_j^{(0)}}
\nuc{\qw}{q^{(1)}}
\nuc{\Ntilu}{\tilde N_u}
\nuc{\Ntilt}{\tilde N_t}
\nuc{\theh}{\theta^{(h)}}
\nuc{\thew}{\theta^{(1)}}
\nuc{\thewT}{\theta^{(1)\top}}

\nuc{\Thew}{\Theta^{(1)}}

\nuc{\NtilT}{\tilde N_T}
\nuc{\RTh}{R_T^{(h)}}
\nuc{\RTo}{R_T^{(0)}}
\nuc{\RTw}{R_T^{(1)}}

\nuc{\chihTh}{\hat\chi_T^{h}}
\nuc{\chihThT}{\hat\chi_T^{h\top}}
\nuc{\chihTw}{\hat\chi_T^{(1)}}
\nuc{\chihTwT}{\hat\chi_T^{(1)\top}}

\nuc{\STh}{S_T^{(h)}}
\nuc{\STw}{S_T^{(1)}}
\nuc{\STo}{S_T^{(0)}}

\nuc{\Nh}{\hat N_T}
\nuc{\GTh}{G_T^{(h)}}
\nuc{\GTo}{G_T^{(0)}}
\nuc{\GTw}{G_T^{(1)}}

\nuc{\gTw}{g_T^{(1)}}

\nuc{\sTh}{s_T^{(h)}}
\nuc{\sTw}{s_T^{(1)}}
\nuc{\sTwT}{s_T^{(1)\top}}
\nuc{\sTo}{s_T^{(0)}}

\nuc{\LamhT}{\hat\Lam_T}

\nuc{\Lamh}{\hat\Lambda}
\nuc{\lamh}{\lam^{(h)}}
\nuc{\chijh}{\chi_j^{(h)}}
\nuc{\chio}{\chi^{(0)}}
\nuc{\RTho}{R_T^{(h,0)}}
\nuc{\alpo}{\alpha^{(0)}}
\renuc{\Rsh}{R_*^{(h)}}
\nuc{\Rsw}{R_*^{(1)}}
\nuc{\Rso}{R_*^{(0)}}
\nuc{\Rsho}{R_*^{(h,0)}}
\nuc{\Rswo}{R_*^{(1,0)}}
\nuc{\muh}{\mu^{(h)}}
\nuc{\muhT}{\mu^{(h)\top}}
\nuc{\muo}{\mu^{(0)}}
\nuc{\muoT}{\mu^{(0)\top}}
\nuc{\muw}{\mu_1}
\nuc{\muwT}{\mu_1^{\top}}
\nuc{\qh}{q^{(h)}}
\nuc{\qbh}{\bar q^{(h)}}
\nuc{\qbo}{\bar q^{(0)}}
\nuc{\qo}{q^{(0)}}
\nuc{\ssh}{s_*^{(h)}}
\nuc{\Nv}{N_v}

\nuc{\lamo}{\lam^{(0)}}
\nuc{\lamw}{\lam^{(1)}}
\nuc{\alpjo}{\alpha_j^{(0)}}
\nuc{\alpjw}{\alpha_j^{(1)}}
\nuc{\betw}{\beta^{(1)}}

\nuc{\Lto}{L_2[0,\infty)}
\nuc{\gw}{g^{(1)}}
\nuc{\gbw}{\bar g^{(1)}}

\nuc{\chihT}{\hat\chi_T}
\nuc{\alphT}{\hat\alpha^{(h)\top}}
\nuc{\alphTw}{\hat\alpha_T^{(1)}}
\nuc{\chTw}{\hat c_T^{(1)}}

\nuc{\wnPw}{\bm{1}_{P^{(1)}}}
\nuc{\wnPh}{\bm{1}_{P^{(h)}}}
\nuc{\alpsh}{\alpha_*^{h}}
\nuc{\ssw}{s_*^{(1)}}
\nuc{\VhTh}{\hat V_T^{(h)}}
\nuc{\VhTw}{\hat V_T^{(1)}}
\nuc{\RTwo}{R_T^{(1,0)}}

\nuc{\Sigh}{\Sigma^{(h)}}
\nuc{\Sigw}{\Sigma^{(1)}}
\nuc{\Sigo}{\Sigma^{(0)}}

\nuc{\phio}{\phi^{(0)}}
\nuc{\phiw}{\phi^{(1)}}
\nuc{\Dphi}{\Delta\phi}
\nuc{\RThh}{R_T^{(h,h)}}
\nuc{\chihp}{\chi^{(h')}}
\nuc{\chitilhp}{\tilde\chi^{(h')}}
\nuc{\phisw}{\phi^{(1)}_*}
\nuc{\Gamsw}{\Gam_*^{(1)}}
\nuc{\Gamsh}{\Gam_*^{(h)}}
\nuc{\Gamso}{\Gam_*^{(0)}}
\nuc{\Fphih}{F_\phi^{(h)}}
\nuc{\Fphiw}{F_\phi^{(1)}}
\nuc{\Fphio}{F_\phi^{(0)}}
\nuc{\Sphih}{S_\phi^{(h)}}
\nuc{\Sphiw}{S_\phi^{(1)}}
\nuc{\Sphio}{S_\phi^{(0)}}
\nuc{\phih}{\phi^{(h)}}
\nuc{\sqintT}{\wo{\sqrt{T}}\intoT}
\nuc{\intT}{\wo{T}\intoT}
\nuc{\intoTs}{\int_{(0,T]^2}}
\nuc{\phibw}{\bar\phi^{(1)}}
\nuc{\phibws}{\bar\phi^{(1)}_*}

\dmo{\WT}{W_{j}}
\nuc{\WsT}{W^*_T}
\nuc{\intoum}{\int_0^{u_-}}
\nuc{\qfpf}{q\star f\cdot \Dphi\star f}
\nuc{\qfpfu}{q\star f(u)\Dphi\star f(u)}

\nuc{\intTi}{\int_T^\infty}
\nuc{\intui}{\int_u^\infty}
\nuc{\intotT}{\int_0^{\tau T}}
\nuc{\intoTmu}{\int_0^{T-u}}

\nuc{\Sef}{S_{ef}}

\renuc{\lamw}{\lam_1}
\renuc{\alpw}{\alpha_1}
\nuc{\chiw}{\chi_1}
\nuc{\Lw}{L^1[0,\infty)}
\renuc{\Lt}{L^2[0,\infty)}
\nuc{\Lwp}{L^1_+[0,\infty)}
\nuc{\Lwn}[1]{\Ver{#1}_{L^1}}
\nuc{\Lin}[1]{\Ver{#1}_{L^\infty}}
\nuc{\Lpn}[1]{\Ver{#1}_{L^p}}
\nuc{\Ltn}[1]{\Ver{#1}_{L^2}}
\renuc{\Li}{L^\infty[0,\infty)}
\nuc{\lamtil}{\tilde\lam}
\nuc{\chitil}{\tilde\chi}
\renuc{\phio}{\phi_0}
\renuc{\phiw}{\phi_1}
\renuc{\lamo}{\lam_0}
\renuc{\lamtilw}{\tilde\lam_1}

\nuc{\intotm}{\int_0^{t-}}
\nuc{\intium}{\int_{-\infty}^{u-}}
\nuc{\intiu}{\int_{-\infty}^{u}}
\nuc{\theT}{\theta^\top}
\nuc{\chihTT}{\hat\chi_T^\top}
\renuc{\chihT}{\hat\chi_{_T}}
\renuc{\chihTT}{\hat\chi_T^\top}
\renuc{\chihTw}{\hat\chi_{_T}}
\renuc{\chihTwT}{\hat\chi_T^\top}
\nuc{\GT}{G_T}
\nuc{\gT}{g_{_T}}
\nuc{\RT}{R_T}
\nuc{\sT}{s_{_T}}
\renuc{\ch}{\hat c}
\renuc{\theh}{\hat\theta}
\renuc{\chio}{\chi_0}
\renuc{\alpo}{\alp_0}
\nuc{\bw}{\mathbbm{1}}
\nuc{\Rs}{R_*}
\nuc{\muT}{\mu^\top}
\nuc{\qb}{\bar q}
\renuc{\phibo}{\bar\phi_0}
\nuc{\alps}{\alpha_*}
\nuc{\cs}{c_*}
\renuc{\co}{c_0}

\nuc{\Creg}{C_{reg}}
\nuc{\kap}{\kappa}
\nuc{\kapc}{\check\psi}
\nuc{\intiom}{\int_{-\infty}^{0-}}

\nuc{\snwi}{\sum_{n=1}^\infty}

\nuc{\chijp}{\chi_{j'}}
\nuc{\chitilj}{\tilde\chi_j}
\nuc{\intoTm}{\int_0^{T-}}

\nuc{\VhT}{\hat V_T}
\nuc{\thes}{\theta_*}
\renuc{\chitilo}{\tilde\chi_0}
\nuc{\Dchio}{\Delta\chi_0}
\nuc{\kapb}{\bar\kap}

\nuc{\BaT}{B^\alp_T}
\nuc{\BcT}{B^c_T}
\nuc{\ot}{^{\otimes 2}}

\nuc{\Sigoth}{\Sig_0^\theta}
\nuc{\Sigoalp}{\Sig_0^\alpha}
\nuc{\Sigcrb}{\Sig_{CRB}}
\nuc{\sigoc}{\sig_0^c}
\renuc{\Sigo}{\Sig_0}
\nuc{\Gs}{G_*}

\nuc{\xitil}{\tilde\xi}
\nuc{\Dchi}{\Delta\chi}
\nuc{\Dchij}{\Delta\chi_j}

\nuc{\Gams}{\Gam_*}
\nuc{\pt}{pseudo-true}
\nuc{\Mr}{Martingale representation}
\nuc{\mr}{martingale representation}
\nuc{\phis}{\phi_*}
\nuc{\phij}{\phi_j}

\dmo{\Sr}{S}
\nuc{\Sqj}{S^q_j}
\nuc{\Sq}{S^q}
\renuc{\Sphio}{S^\phi_0}
\nuc{\Sphis}{S^\phi_*}
\nuc{\Skap}{S^\kap}
\nuc{\sqT}{\sqrt{T}}
\nuc{\phibs}{\bar\phi_*}

\nuc{\DW}{\Delta W_T}
\nuc{\DWj}{\Delta W_{j,T}}
\nuc{\BDGi}{Burkholder-Davis-Gundy inequality}

\nuc{\Vs}{V^{T,u}_s}
\nuc{\VTu}{V^{T,u}}
\nuc{\intosum}{\int_0^{su-}}
\nuc{\intosu}{\int_0^{su}}

\nuc{\Wtil}{\tilde W}
\renuc{\ha}{h^\alp}
\nuc{\Dphio}{\Delta\chi_0}

\renuc{\muh}{\mu_h}
\nuc{\muha}{\mu_{h}^{\alpha}}
\nuc{\xitilh}{\tilde\xi_h}
\renuc{\chitilh}{\chitil_h}

\nuc{\Sigsth}{\Sigma_*^\theta}
\nuc{\Sigsalp}{\Sigma_*^\alp}
\nuc{\sigsc}{\sig_*^c}
\nuc{\Sigs}{\Sig_*}

\nuc{\chihh}{\chi_h}
\nuc{\chis}{\chi_*}

\nuc{\chitilq}{\chitil_q}
\nuc{\phiq}{\phi_q}
\nuc{\lamtilq}{\lamtil_q}
\renuc{\lamtilh}{\lamtil_h}
\nuc{\lamtilg}{\lamtil_g}
\renuc{\phih}{\phi_h}

\nuc{\chitilha}{\tilde\chi_h^\alp}

\nuc{\MtilxihT}{\tilde M^\xi_{h,T}}
\nuc{\MtilxiT}{\tilde M^\xi_{T}}

\nuc{\dome}{\delta_\omega}
\nuc{\dt}{\delta_t}
\nuc{\Dphib}{\overline{\Dphi}}

\renuc{\pro}[1]{{\it Proof. }#1\hfill$\square$}
\nuc{\bSp}{\mathbb{S}_+}

\nuc{\RqT}{R_{T}^q}
\nuc{\RgT}{R_{T}^g}
\nuc{\RhT}{R_{T}^h}
\renuc{\sqT}{s_{T}^q}
\nuc{\sgT}{s_{T}^g}
\nuc{\shT}{s_{T}^h}

\nuc{\chihqT}{\hat\chi_{T}^q}
\nuc{\chihqTT}{\hat\chi_T^{q\top}}
\nuc{\chiq}{\chi_q}
\nuc{\Rsqo}{R_*^{(q,0)}}
\nuc{\Rsgo}{R_*^{(g,0)}}
\nuc{\muq}{\mu_q}
\renuc{\muh}{\mu_h}
\nuc{\Rsq}{R_*^q}
\nuc{\Rsg}{R_*^g}
\renuc{\Rsh}{R_*^h}
\nuc{\csq}{c_*^q}
\nuc{\alpsq}{\alp_*^q}
\nuc{\Gamsq}{\Gam_*^q}
\nuc{\alphTq}{\hat\alp_T^q}
\nuc{\chTq}{\hat c_T^q}

\nuc{\alphTh}{\alph_T^h}
\nuc{\alphTg}{\alph_T^g}
\nuc{\alphThT}{\alph_T^{h\top}}
\nuc{\chTh}{\hat c_T^h}
\nuc{\shTT}{s_T^{h\top}}
\nuc{\alpsqT}{\alp_*^{q\top}}
\nuc{\alpshT}{\alp_*^{h\top}}

\renuc{\clrb}{}

\title{\LARGE \bf
A System-Theoretic Approach to Hawkes Process Identification with Guaranteed Positivity and Stability}

\author{Xinhui Rong and Girish N. Nair
\thanks{X. Rong ({\tt\small xinhui.rong@unimelb.edu.au}) and G. N. Nair {\tt\small gnair@unimelb.edu.au}) are with the Department of Electrical and Electronic Engineering, University of Melbourne, Australia. 
}
}%

\begin{document}

\renewcommand{\theequation}{\arabic{section}.\arabic{equation}}

\newtheorem{theorem}{Theorem}
\newtheorem{lemma}{Lemma}
\newtheorem{result}{Result}
\newtheorem{corollary}{Corollary}[theorem]
\newtheorem{proposition}{Proposition}[theorem]
\newtheorem{thm}{Theorem}
\newtheorem{lem}{Lemma}
\newtheorem{rst}{Result}
\newtheorem{defn}{Definition}
\newtheorem{remark}{Remark}
\newtheorem{coro}{Corollary}[thm]
\newtheorem{prop}{Proposition}[thm]

\maketitle
\thispagestyle{empty}
\pagestyle{empty}

\begin{abstract}
The Hawkes process models self-exciting event streams, requiring a strictly non-negative and stable stochastic intensity. Standard identification methods enforce these properties using non-negative causal bases, yielding conservative parameter constraints and severely ill-conditioned least-squares Gram matrices at higher model orders. To overcome this, we introduce a system-theoretic identification framework utilizing the sign-indefinite orthonormal Laguerre basis, which guarantees a well-conditioned asymptotic Gram matrix independent of model order. We formulate a constrained least-squares problem enforcing the necessary and sufficient conditions for positivity and stability. By constructing the empirical Gram matrix via a Lyapunov equation and representing the constraints through a sum-of-squares trace equivalence, the proposed estimator is efficiently computed via semidefinite programming.
\end{abstract}

\mysec{Introduction}\label{sec:intro}
In recent years, the unprecedented availability of asynchronous random event data, 
arising from fields as diverse as computational neuroscience \cite{Bial97}, 
genomics \cite{Cars10} and 
high-frequency finance \cite{Bacr15}, 
has driven a surge in neuromorphic technologies across modern robotics \cite{Gall22} and control systems \cite{Shi15}. 
The Hawkes process is widely regarded as the standard model for characterizing the history-dependent dynamics inherent in these event-driven systems. 

A Hawkes process is defined by an intensity function that characterizes the instantaneous event rate. 
The standard maximum likelihood estimation (MLE) \cite{Ozak79}, computed via direct gradient descent, suffers from non-convexity caused by nonlinear parameters in the Hawkes impulse response (HIR).
Therefore, a standard approach is to approximate the HIR 
using a dictionary of prescribed/user-defined causal basis functions, yielding a linear model. 
This facilitates the use of expectation-maximization (EM) algorithms 
\cite{Godo20, Lewi11, Veen08} with explicit iterations and 
enables least-squares (LS) algorithms 
\cite{Bacr20, Hans15, Rong26} 
which allow for parameter constraints.

Choosing suitable basis functions is a non-trivial task. 
The orthonormal Laguerre basis \cite{Wahl91} 
is theoretically favored for Hawkes processes \cite{Godo20, Ogat82} 
due to its density \cite{Szeg59} in both $\Lw$ and $\Lt$.
Yet, the use of the Laguerre basis in Hawkes identification is hindered by its computational complexity and the critical requirement of a non-negative intensity \clrb{\cite{Solo24}}. 
Consequently, the Erlang basis \cite{Godo20} is adopted as a convenient positive alternative with a tractable formulation and a guaranteed positivity through positive parameter initialization in EM algorithms \cite{Godo20, Lewi11}.

However, this loss of orthogonality incurs a severe numerical penalty in LS estimation. 
\clrb{For discrete-time linear systems, it is known that under non-orthonormal bases, increasing the model order drives the Gram matrix towards singularity \cite{Wahl91}. 
Our recent work \cite{Rong26a} empirically demonstrates a similar phenomenon for point processes under the Hawkes-Erlang bases.}
This ill-conditioning causes standard LS algorithms to fail and the asymptotic parameter error covariance to diverge. 
Moreover, like many other positive bases \cite{Xu16, Zhou13}, 
enforcing the non-negativity of the intensity function 
by demanding non-negative parameters is merely a sufficient condition.  
This approach becomes increasingly conservative 
as the model order grows, excluding a vast subspace of physically valid Hawkes models.  

To overcome these limitations, we model the HIR via the sign-indefinite orthonormal Laguerre basis and formulate a constrained LS estimation framework. 
Our system-theoretic contributions are threefold. 
\clrb{
(i) Extending Wahlberg’s condition-number analysis from discrete-time linear systems \cite{Wahl91} to point processes, we establish the well-conditioning of the asymptotic Gram matrix of the proposed OrthoNormal Hawkes-Laguerre (ON-HL) model, independent of the model order.
We then develop the full toolkit needed to identify an ON-HL model.
(ii) We repurpose the classical continuous-time state-space realization \cite{Solo12} in a new LS setting to efficiently compute the empirical Gram matrix in continuous time via a Lyapunov equation, which is, to our knowledge, new. 
(iii) We provide the first sum-of-squares (SOS) positivity constraint for Hawkes processes and introduce a novel trace-based equivalence that yields a convex semidefinite program (SDP) enforcing both positivity and stability.}

We organize the paper as follows. Section \ref{sec:pre} reviews the Hawkes process preliminaries and Laguerre polynomial properties. Section \ref{sec:ls} formulates an unconstrained centered LS problem for ON-HL identification, providing 
\clrb{the condition-number bounds and the efficient state-space-based computations.} 
Section \ref{sec:cls} formulates the SDP problem for a positive and stable ON-HL model. Section \ref{sec:sim} contains comparative simulations, and Section \ref{sec:con} concludes the paper.

\mysec{Preliminaries} \label{sec:pre}
We review the properties of the Hawkes process  
and the Laguerre polynomials. 
We provide the theoretical basis 
of the synthesis of point process statistics and systems theory. 

\subsection{Hawkes Processes}
The Hawkes process \cite{Hawk71} describes history-dependent 
self-exciting event streams, a.k.a. {\it point processes} \cite{Dale03}. 
A point process is equivalent to a {\it counting process} 
$N_t \trieq N((-\infty, t])$ where $N(A)$ counts the number 
of events that occur in the time set $A\subset\bR$. 
$N_t$ is a non-decreasing step function that jumps by one 
at each random event time $T_r$. 
The standard {\it orderliness} \cite{Dale03} condition $\limdo\frac{\Pr[N([t,t+\delta))>1]}{\Pr[N([t,t+\delta))=1]}=0$ is in place throughour the paper.

The Hawkes process is characterized by a (conditional) {\it intensity function} 
defined as $\lamo(t) \trieq	\limdo\wo\delta\Pr[N([t,t+\delta))=1|\cHtm]$ 
with $\cHtm$ being the history/filtration $\sigma\{N_s:s<t\}$ 
and the {\it Hawkes intensity} is given by
\eqn{
\lamo(t) =  c_0 + \chio(t)	= c_0 + \intitm\phio(t-u)dN_u,\label{eq:lam}
}
which is itself stochastic, describing 
the limiting probability rate of an increase in the counting process $N$ 
conditioned on the full history $\cHtm$, 
and consists of a constant {\it background rate} $c_0$ 
and a stochastic {\it memory process} $\chio(t)\trieq\intitm\phio(t-u)dN_u=\sum_{T_r<t} \phio(t-T_r)$, where $\phio:[0,\infty)\mapsto[0,\infty)$ is the deterministic and causal 
{\it Hawkes impulse response (HIR)}. 

The {\it stationarity/stability} condition for a Hawkes process is that 
the {\it branching ratio} $\Gam\trieq\intoi\phio(t)dt<1$. 
Under the condition $\Gam<1$, there exists a unique strictly stationary distribution \cite{Brem96}
for the counting process $N_t$ characterized by the intensity \eqref{eq:lam} 
and the {\it expected rate} $\E[dN_t]/dt=\E[\lamo(t)]=\E[\lamo(0)]\trieq\Lam=\frac{c_0}{1-\Gam}$. Throughout the paper, the expectation $\E$ 
is taken with respect to such a unique stationary probability measure.

\subsection{Laguerre Bases}
The Laguerre basis functions $h_j(t)$ are conveniently defined by their Laplace transforms (LTs) \cite{Wahl91}
\eqn{\label{eq:h}
\bar h_j(s) = \frac{\sqrt{2\beta}}{s+\beta}\bra{\frac{\beta-s}{\beta+s}}^j, \quad j=0,1,2,\dotsm.
}
Throughout the paper, we denote the LT of a function $f:[0,\infty)\mapsto\bR$ by $\bar f(s) = \intoi e^{-st}f(t)dt$. 
The definition in \eqref{eq:h} has a sign difference from the original definition in \cite{Wahl91} at odd orders. This is to ensure that each $\intoi h_j(t)dt = \bar h_j(0)=\sqrt{2/\beta}$ for regularization convenience. 

Recovering $h_j(t)$ in the time domain, 
we have $h_j(t) = w(t) u_j(t)$, where the weight function
$w(t) = \beta e^{-\beta t}\cI_{t\geq0}$ 
and the Laguerre polynomials $u_{j}(t) = \sum_{0\leq k\leq j} \frac{\sqrt{2/\beta}(-1)^{j-k}\binom{j}{k}(2\beta t)^k}{k!}\cI_{t\geq0}$, with $\cI_{t\in A}=\{\smat{1& t\in A\\0&t\notin A}$ as the indicator function. 
The Laguerre polynomials and the Laguerre basis functions 
are orthonormal in the sense that 
\eq{
	&\intoi u_j(t)u_{j'}(t)w^2(t)dt 
=	\intoi h_j(t)h_{j'}(t)dt= \cI_{j=j'}.
}

Orthonormal polynomials satisfy a three-term recursion \cite{Roh06} that is key to our development. We adopt the notation $a_j^k \trieq [a_j,a_{j+1},\dotsm,a_k]^\top, k>j$ to represent a stacked vector. We also define $\bSp^k$ to be the space of real and symmetric positive definite matrices of dimension $k\times k$. 
\Blem{lem:3tr} \cite{Roh06} 
The orthonormal Laguerre polynomials $u_j(t)$ with 
$w(t)=\beta e^{-\beta t}\cI_{t\geq0}$, satisty a three-term recursion
\eqn{
u_{j+1}(t) = (\rho_j t-\kap_j)u_j(t) - \gam_j u_{j-1}(t),\label{eq:3tr}
}
with $\rho_j = \frac{2\beta}{j+1}, \kap_j=\frac{2j+1}{j+1}, \gam_j=\frac{j}{j+1}, u_0(t)=\sqrt{2/\beta}$, 
and $u_j(t)\trieq0$ for $j<0$. As a property, $\wo{\rho_{j}}=\frac{\gam_{j+1}}{\rho_{j+1}}$. For an arbitrary dimension $m\geq1$, the three-term recursion takes a vector form
$t u_0^{m-1}(t) = J_m u_0^{m-1}(t) + \wo{\rho_{m-1}}u_m(t)e_m$, with $e_m=[0,\dots,0,1]^\top\in\bR^m$ and the symmetric tridiagonal Jacobi matrix 
$J_m\in\bSp^m$ 
which contains $\frac{\kap_0}{\rho_0},\dotsm,\frac{\kap_{m-1}}{\rho_{m-1}}$ 
down its main diagonal and $\wo{\rho_0},\dotsm,\wo{\rho_{m-2}}$ down 
its immediate upper and lower subdiagonals.
\Elem

\mysec{Unconstrained LS for ON-HL} \label{sec:ls}
This section establishes the unconstrained LS estimation using the ON-HL representation to approximate \eqref{eq:lam}. We theoretically show that, unlike the severely ill-conditioned Hawkes-Erlang alternative, the ON-HL asymptotic Gram matrix is inherently well-conditioned at an arbitrary model order. Finally, we derive an efficient system-theoretic computation for the empirical Gram matrix and its associated regressors. We assume the following standard assumptions. 

\Bass{A1} {\it Stationarity. }
The true counting process $N_t$ admits the unique 
stationary probability measure \cite{Brem96} 
associated with the intensity \eqref{eq:lam} with $\Gam<1$. 
The observed counting process $\Ntilt \trieq N((0,t])$ 
is a truncation of $N_t$ within the deterministic observation period $t\in[0,T]$ 
and $\NtilT\geq1$.  
\Eass

\Bass{A2}{\it HIR regularization. }
The true HIR $\phio$ satisfies (a) $\phio\in\Lw\cap\Li$, and (b) $\intoi t\phio(t)dt<\infty$. 
\Eass

The assumptions \ref{A1} and \ref{A2} are standard in the literature for asymptotic analysis \cite{Eich17,Brem96}. 
Under \ref{A1}, the increments $d\Ntilt=\cI_{t>0}d\Nt$ 
remain strictly stationary and ergodic \cite[Ch.12]{Dale08}. 
\ref{A2} ensures an almost-surely bounded intensity function  
and the asymptotic vanishing of the transients originating from the unobserved pre-sample history \cite{Brem96,Rong26a}.  
\ref{A2} is broadly satisfied, including mixtures of exponentials 
and heavy-tailed defective densities with a finite mean.

\subsection{The Truncated Hawkes Intensity}
A truncated intensity with a causal basis $q$ is defined as
\eqn{
&\lamtilq(t) = c + \alpha^\top\chitilq(t)
	=c + \intotm \phiq(t-u;\alpha)d\Ntilu \label{eq:lam1}\\
&\phiq(t;\alpha) =\alpha^\top q(t)\geq0\label{eq:phi1}\\
&\chitilq(t) = \intotm q(t-u)d\Ntilu\in\bR^P\label{eq:chi},
}
where 
$\lamtilq(t)$ is the $P$-th order truncated intensity parameterized by 
the candidate background rate $c>0$ and the candidate basis weights $\alpha\in\bR^P$. 
The candidate HIR $\phiq:[0,\infty)\mapsto[0,\infty)$ is constructed by the prescribed 
causal basis $q(t) = q_0^{P-1}(t)\in\bR^P$,
and $\chitilq$ is the stochastic {\it truncated memory regressor}. 
The events before time $0$ are not observed.

We propose using the ON-HL model to approximate an arbitrary Hawkes intensity 
\eqref{eq:lam} under \ref{A1}, \ref{A2}. 
Define the Laguerre basis vector $h(t)=h_0^{P-1}(t)=w(t)u(t)\in\bR^P$ 
with $u(t)=u_0^{P-1}(t)\in\bR^p$. 
The truncated ON-HL intensity $\lamtilh(t)$ is defined through \eqref{eq:lam1}-\eqref{eq:chi} by setting $q=h$.

In parallel, the truncated Hawkes-Erlang intensity $\lamtilg(t)$, 
currently used in the literature, is defined by
replacing $q$ in \eqref{eq:lam1}-\eqref{eq:chi} by $g$ with 
$g(t) = g_0^{P-1}(t)=w(t)v(t)\in\bR^P_{\geq0}$ where $g_j(t) = w(t)v_j(t)$ 
is the Erlang density with
$v_j(t) = \frac{(\beta t)^{j}}{j!}$ and $v(t)=v_0^{P-1}(t)$. 
The Erlang basis $g(t)=Lh(t)$ is linearly transformed from the Laguerre basis $h$ by a lower triangular  matrix 
$L$ with $L_{ij}=\sqrt{\beta/2}{\choose{i-1}{j-1}}/{2^{i-1}}\cI_{j\leq i}$ and 
$(L^{-1})_{ij}=\sqrt{2/\beta}(-1)^{i-j}2^{j-1}\choose{i-1}{j-1}\cI_{j\leq i}$ as the $(i,j)$-th component of  $L$ and $L^{-1}$, respectively.

\subsection{\clrb{The Unconstrained LS Estimator}}
\clrb{The LS framework is based the martingale increment \cite{Brem81} property of 
$dN_t - \lamo(t)dt$.  
Working with the truncated observation $\tilde N$ and the basis approximation $\lamtilq$, 
the ordinary LS objective is given by  \cite{Hans15}  $\haf\intoT \lamtil_q(t)^2dt - \intoT\lamtil_q(t)d\Ntilt$. 
To decouple $\alpha$ from the background rate $c$, 
we work with an equivalent \cite{Rong26} centered LS (CLS)
which sets\footnote{\clrb{Heuristically, $\NtilT\approx\intoT\lamtilq(t;\alp)dt\Ra c\approx \frac{\NtilT}T-\woT\alp^\top\chitilq(t)dt$.}} 
$c=\frac{\NtilT}T-\woT\intoT\alpha^\top\chitilq(t)dt$ 
and the objective becomes}
\eqn{\label{eq:J}
	J_{T}^q(\alp)
=	&\haf\alp^\top \RqT\alp - \alp^\top \sqT, 
}
where $\sqT=\woT\intoT\chitilq(t)d\Ntilt-\LamhT\chihqT\in\bR^P$ 
is the empirical cross-covariance 
with $\LamhT=\frac{\NtilT}{T}$ and $\chihqT = \woT\intoT\chitilq(t)dt$, 
and the {\it empirical Gram matrix} is an empirical covariance
\eqn{\label{eq:R}
\RqT = \woT\intoT\chitilq(t)\chitilq(t)^\top dt - \chihqT\chihqTT\in\bR^{P\times P}.
}
Given that $\RqT$ is positive definite, the LS estimators are 
\eqn{
\alphTq &= \argmin_{\alpha\in\bR^P}J_T^q(\alp) = (\RqT)^{-1}\sqT\label{eq:alph}\\
\chTq &= \LamhT-\chihqTT\alphTq,\label{eq:ch}
}
where \eqref{eq:alph} follows directly from matrix calculus solving $\frac\partial{\partial\alpha}J_T^q(\alp)=0$ and \eqref{eq:ch} recovers the centering.

\subsection{\clrb{The Eigenvalue Bounds on the Asymptotic Gram Matrices}}
\clrb{In \cite{Wahl91}, orthonormal Laguerre bases are shown to yield well-conditioned asymptotic Gram matrices for discrete-time linear systems. Here, we establish the analogous result for Hawkes LS identification. This uses the following lemma, which extends the ergodic lemma in \cite{Rong26a} by relaxing the positivity and unit-mass assumptions on the kernel basis.}

\Bass{A3} {\it Basis conditions. }
(a) $q_j\in\Lw\cap\Li$, (b) $\intoi t|q_j(t)|dt<\infty$. 
Further, (c) for any $\tau>0$, 
there is no pair of $x\neq0\in\bR^{P}$ and $d\in\bR$, 
such that $x^\top q(t)= d$, almost everywhere on $[0,\tau]$. 
\Eass

\clrb{\Blem{lem:asym}
Suppose \ref{A1}-\ref{A3} are satisfied. 

(a) $\RqT>0$ with probability one (w.p.1), so \eqref{eq:alph} exists and uniquely minimizes \eqref{eq:J}. 

(b) As $\Ttoi$, w.p.1, 
\eq{
\RqT&\to\Rsq\trieq\E[(\chiq(0)-\muq)(\chiq(0)-\muq)^\top]\\
	&= \wo{2\pi}\intii \bar q(\jw)\bar C(\ome)\bar q(-\jw)^\top d\ome\in\bSp^P,
}
where $\Rsq$ is the {\it asymptotic Gram matrix}, $\chiq(t) = \intitm q(t-u)dN_u$ is the stationary full memory regressor with $\muq \trieq \E[\chiq(t)]=\Lam \intoi q(t)dt$, 
and 
$
\bar C(\ome)=\frac{\Lam}{|1-\phibo(\jw)|^2},
$
is Bartlett's spectrum \cite{Hawk71}, 
which is the Fourier transform of 
the covariance density $C(u-v) = \frac{\E[dN_udN_v]-\Lam^2}{dudv}$. 
\Elem}

\clrb{The analogous definitions for the ON-HL and Hawkes-Erlang models respectively replace $q$ with $h$ and $g$. 
Under \ref{A1} and \ref{A2}, Lemma \ref{lem:asym} holds for both the Laguerre and Erlang bases, since they trivially satisfy \ref{A3}. 
The asymptotic convergence of the LS estimators $\alphTq,\chTq$ 
follows by extending \cite[Theorem 15]{Rong26a} to sign-indefinite bases. We omit these results here and focus on the identification problems for ON-HL.
}

The following results show that the ON-HL model is superior to the Hawkes-Erlang model in the LS setting, regarding the well-conditioning of the asymptotic Gram matrix. 
We denote the smallest and largest absolute eigenvalues of a matrix $A$ by $\rho_{\min}(A)$ and $\rho_{\max}(A)$, and the smallest and largest singular values by $\sig_{\min}(A)$ and $\sig_{\max}(A)$, respectively.

\Bthm{thm:eigR}
Under \ref{A1} and \ref{A2}, the asymptotic Gram matrix $\Rsh$ under the Laguerre basis $h$ satisfies
\eq{
&\rho_{\min}(\Rsh)\geq \frac{\Lam}{(1+\Gam)^2},\quad
\rho_{\max}(\Rsh)\leq \frac{\Lam}{(1-\Gam)^2}, 
}
whereas under the Erlang basis $g$, the asymptotic Gram matrix $\Rsg$ satisfies 
$\Rsg = L\Rsh L^\top$, and
\eq{
&\rho_{\min}(\Rsg)\leq\rho_{\max}(\Rsh)\sig_{\min}^2(L)\leq\frac{\beta\Lam}{2(1-\Gam)^2}4^{-(P-1)}\\
&\rho_{\max}(\Rsg)\geq\rho_{\min}(\Rsh)\sig_{\max}^2(L)\geq\frac{\beta\Lam}{2(1+\Gam)^2}.
}
\Ethm

The condition number of the ON-HL Gram matrix is upper-bounded by $\frac{(1+\Gam)^2}{(1-\Gam)^2}$,  independent of $P$. Conversely, the Hawkes-Erlang condition number is lower-bounded by $\frac{(1-\Gam)^2}{(1+\Gam)^2}4^{P-1}$, indicating exponential ill-conditioning. 
Finite-precision calculations will frequently render the Erlang matrix $\RgT$ indefinite at high orders. 
The conditioning of both $\Rsh$ and $\Rsg$ inevitably worsens as $\Gam$ approaches $1$, reflecting a system near instability. 

\subsec{System-Theoretic Computation of the Estimators}
\clrb{Here, we adopt a system-theoretic approach, 
computing $\RhT$ exactly by solving a Lyapunov equation for ON-HL, 
thereby avoiding the cumbersome piecewise-constant approximations \cite{Hans15} and entry-wise recursions \cite{Bacr20} that can degrade conditioning. 
Our method builds on the standard state-space representation, 
previously used for maximum-likelihood estimation \cite{Solo12,Gere21}, and uses it, for the first time, to compute the empirical Gram matrix in the LS setting.

\Blem{lem:ss} 
Under ON-HL, $\chitilh$ 
satisfies the stochastic differential equation \cite{Solo12}
\eqn{
d\chitilh(t) = A\chitilh(t)dt + Bd\Ntilt,\label{eq:sde}
}
where $A\in\bR^{P\times P}$ 
is lower-triangular with $-\beta$ on the main diagonal 
and $(-1)^{i+1}2\beta$ on each $i$-th lower subdiagonal, and 
$B\in\bR^{P}$ has $(-1)^{i+1}\sqrt{2\beta}$ at its $i$-th component. 
\Elem

Given an observation with  $\NtilT=n$ 
and the increasing event times 
$[T_1,\dotsm,T_{\NtilT}]^\top= [t_1,\dotsm,t_n]^\top$, 
the computation of the first-order terms $\shT, \chihqT$ are straightforward by 
the following vector recursions.
\blist{
\item $\chitilh(t_1) = 0, \chitilh(t_{(r-1)+}) = \chitilh(t_{r-1})+B, \chitilh(t_r) = e^{A(t_r-t_{r-1})}\chitilh(t_{(r-1)+}), 2\leq r\leq n$.
\item $\chitilh(T) = e^{A(T-t_n)}\chitilh(t_{n+}), \chihTh=\woT A^{-1}(\chitilh(T)-\NtilT B), \LamhT=n/T, \shT=\woT\srwn \chitilh(t_r) - \LamhT\chihTh$.
}

\Bthm{thm:cal} $\RhT$ solves the Lyapunov equation
\eq{
A\RhT+\RhT A^\top + \LamhT BB^\top+B\shTT + \shT B^\top + D_T=0,
}
where $D_T=\woT\chihTh\chihThT - \woT(\chitilh(T) - \chihTh)(\chitilh(T)-\chihTh)^\top$.
\Ethm

The matrix $A$ is Hurwitz by construction, ensuring that the Lyapunov equation yields a unique, real, and symmetric solution \cite{Kail00} $\RhT$. Although the indefiniteness of $D_T$ precludes a direct structural guarantee of positive definiteness from the equation itself, the requisite condition $\RhT > 0$ is inherited from Lemma \ref{lem:asym}(a) and the asymptotic well-conditioning of $\Rsh$ in Theorem \ref{thm:eigR}. 
Lemma \ref{lem:ss} and Theorem \ref{thm:cal} hold analogously for the Erlang basis, where $A$ becomes lower bidiagonal with $-\beta$ on its main diagonal and $\beta$ on its first lower subdiagonal, and $B$ is sparse with $\beta$ solely as its first entry.}

\mysec{Positive and Stable LS for ON-HL} \label{sec:cls}
We formulate the constrained LS for ON-HL to ensure 
that the identified intensity is positive 
($
\phih(t;\alph_T^h)\geq0\mbox{ and }\ch_T^h>0
$)
and stable 
($
\intoi\phih(t;\alph_T^h)dt<1
$). 
However, $\chTh>0$ is indicated\footnote{From \eqref{eq:ch}, we have $\chTh=\LamhT-\chihThT\alphTh$, but
$\chihThT\alphTh=\woT\intoT\intotm \alphThT h(t-u)d\Ntilu dt
=\woT\intoT\intotm\phih(t-u;\alphTh)d\Ntilu dt=\woT\intoT\int_0^{T-u}\phih(t;\alphTh)dtd\Ntilu\leq\woT\intoT\intoi\phih(t;\alphTh)dtd\Ntilu<\LamhT$.}  by 
the other two conditions. 
Therefore, it suffices to restrain the weights $\alp$ 
in the CLS \eqref{eq:J}.  

\clrb{While the stability is enforced by a simple affine inequality, 
non-negativity is guaranteed through its necessary and sufficient condition via the SOS characterization \cite{Roh06}. 
This letter marks the first application of SOS techniques to Hawkes process identification. Moreover, we develop a trace-based SOS reformulation for the Laguerre basis, 
where the associated matrices are not the standard Hankel matrices as in the ordinary polynomial bases (cf. \cite[Ch. 2]{Dumi07}), 
and they can be computed efficiently via simple recursions.
}

\Blem{lem:sos}\cite{Roh06}
Let $f$ be a real polynomial of degree $k$. $f$ is non-negative on $[0,\infty)$, 
iff $f$ can be expressed as an SOS
$
f(t) = a(t)^2 + t b(t)^2$, 
where polynomials $a$ and $b$ have maximum degrees of $\floor{k/2}$ 
and $\floor{(k-1)/2}$, respectively. 
\Elem

\clrb{
\Bthm{thm:sos}
(a) The candidate Laguerre HIR $\phih(t;\alp)=\alp^\top h(t)=w(t)\alp^\top u(t)\geq0$
on $[0,\infty)$ iff there exist $Q_1=Q_1^\top\geq0\in\bR^{m_1\times m_1}, 
Q_2=Q_2^\top\geq0\in\bR^{m_2\times m_2}$ with $m_1=\floor{(P-1)/2}+1, m_2=\floor{P/2}$, such that
\eq{
\alp_j = \trace\{F_jQ_1\} + \trace\{G_jQ_2\}, \quad j=0,1,\dotsm,P-1,
}
where $F_j:=F_{m_1,j}\in\bR^{m_1\times m_1}$ and  $G_j:=G_{m_2,j}\in\bR^{m_2\times m_2}$ satisfy, with the corresponding dimensions, 
\eqn{
u_0^{m-1}(t)u_0^{m-1}(t)^\top& = \ssum{k=0}^{2m-2}F_{m,k}u_k(t),\label{eq:F}\\
tu_0^{m-1}(t)u_0^{m-1}(t)^\top &= \ssum{k=0}^{2m-1}G_{m,k}u_k(t). \label{eq:G}
}
(b) The matrices $F_{m,k}$ and $G_{m,k}$ are the top-left block of 
$F_{M,k}$ and $G_{M,k}$ for any $M>m$ and any $k\leq m$, respectively, and
can be calculated recursively by 
$F_{m,0}=\sqrt{2/\beta}I_m, F_{m,1}=\rho_0(J_m-\frac{\kap_0}{\rho_0}I_m)F_{m,0}$, and 
\eq{
	F_{m,k+1} 
= 	&\rho_k(J_m-\frac{\kap_k}{\rho_k}I_m)F_{m,k}-\frac{\rho_k}{\rho_{k-1}}	F_{m,k-1}\\
	&+\frac{\rho_k}{\rho_{m-1}}\sqbra{\smat{0_{(m-1)\times m}\\F_{m+1,k}^{(m+1)}}}\\
	G_{m,k} 
= 	&\wo{\rho_{k-1}}F_{m,k-1} + \frac{\kap_k}{\rho_k}F_{m,k} + \wo{\rho_k}F_{m,k+1},
}
where $\rho_k,\kappa_k,J_m$ are defined in Lemma \ref{lem:3tr}, 
$I_m\in\bR^{m\times m}$ is an identity matrix, and 
$F_{m+1,k}^{(m+1)}$ is the last row of $F_{m+1,k}$.
\Ethm}

The constrained LS for the ON-HL model enforcing positivity and stability 
is then formulated as an SDP problem
\iary{lcl}{
&\min_{\alp\in\bR^P, Q_1, Q_2} &J_T^h(\alp)=\haf\alp^\top\RhT\alp - \alp^\top\shT\\
&\mbox{subject to } 	&\alp_j = \trace\{F_jQ_1\}+\trace\{G_jQ_2\}, j=0,\dotsm,P-1\\
	&&\ssum{j=0}^{P-1}\alp_j<\sqrt{\beta/2}\\
	&&Q_1=Q_1^\top\geq0,\quad Q_2=Q_2^\top\geq0.
}
\clrb{The convex SDP formulation allows the use of the classical SDP solvers, 
e.g., SDTP3 \cite{SDPT3}, for efficient solutions. 
}

\mysec{Simulations} \label{sec:sim}
In this section, we run comprehensive simulations to compare the proposed ON-HL 
model and the Hawkes-Erlang model under both unconstrained and constrained LS frameworks. 
We design the simulations as follows. 

The true intensity function $\lamo(t)$ has a background rate $c_0=1$ 
and the true HIR $\phio(t) = \sum_{j=1}^3 \alp_{0,j}\beta_{0,j} e^{-\beta_{0,j} t}$ 
with $\alp_{0,1}=\alp_{0,2}=0.3,\alp_{0,3}=0.2$ and $[\beta_{0,1},\beta_{0,2},\beta_{0,3}]^\top=[2,6,16]^\top$. Thus, the true branching ratio $\Gam=0.8$ and the 
expected rate $\Lam = 5$. We consider the observation period $T\in\{50,100,200,400,800\}$ and simulate $M=1000$ Hawkes process trajectories for each $T$ using the thinning algorithm \cite{Ogat81}.  
We consider the model orders $P\in\{5,10,15,20,25\}$ for both models with a common exponent $\beta=5$. 

{\bf Unconstrained LS. }
First, we compare the conditioning of the empirical Gram matrices $\RhT$ and $\RgT$. Under the Laguerre basis, $\RhT$ remains positive definite for all trajectories, upper bounded by a worst condition number of $\frac{\rho_{\max}(\RhT)}{\rho_{\min}(\RhT)} = 106.7$. In contrast, under the Erlang basis, while the matrices $\RgT$ are computed as positive definite for orders $P \le 15$, around $50\%$ and $67\%$ of the matrices at orders $P=20$ and $25$ are identified as indefinite, regardless of the observation period $T$. The best condition numbers for the Hawkes-Erlang model across all realizations scale from $4 \times 10^3$ at $P=5$ to an extreme $6 \times 10^{16}$ at $P=25$. 
We obtain the unconstrained LS estimators $\alphTh=(\RhT)^{-1}\shT$ and $\hat\alp_T^g=(\RgT)^{-1}\sgT$ and 
estimate the asymptotic error covariances $\hat\Sig_q$ ($q=h$ or $g$) by averaging the scaled squared errors $T(\alphTq-\alpsq)(\alphTq-\alpsq)^\top$ across the $M=1000$ realizations at $T=800$. For the ON-HL model, the Frobenius norm ${\Ver{\hat\Sig_h}_F}$ grows modestly from $3.6$ at $P=5$ to $7.0$ at $P=25$. In  contrast, the Hawkes-Erlang norm ${\Ver{\hat\Sig_g}_F}$ explodes from $1.1\times 10^3$ at $P=5$ to $2.8\times 10^{17}$ at $P=20$, ultimately failing at $P=25$. 
We also observe that for both models, the occurrence of unstable estimators increases with the model order, and nearly $100\%$ of the candidate HIRs become sign-indefinite when the order exceeds $P=10$, 
confirming the need for a constrained LS.

\begin{figure}[t]
\hfill
\begin{minipage}[t]{1\linewidth}
\centering{\includegraphics[width=7.5cm]{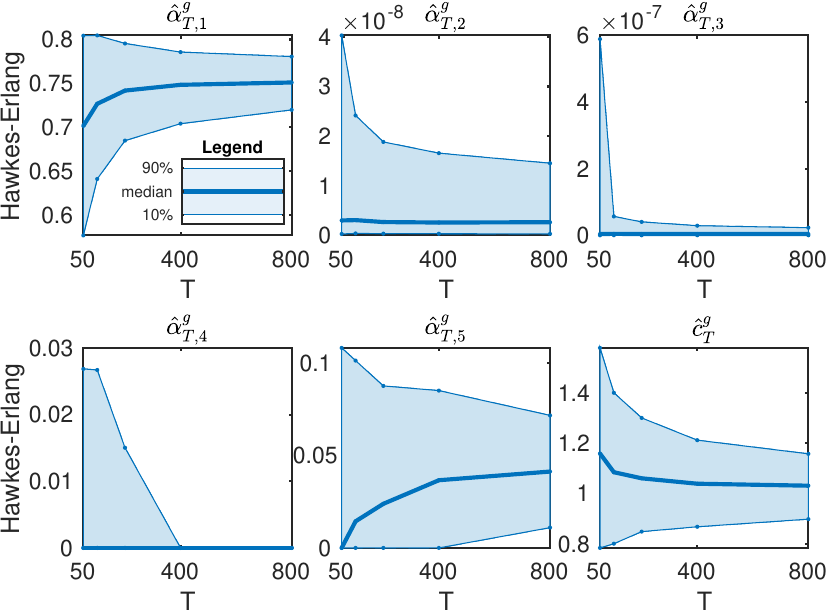}}
\end{minipage}
\begin{minipage}[t]{1\linewidth}
\centering{\includegraphics[width=7.5cm]{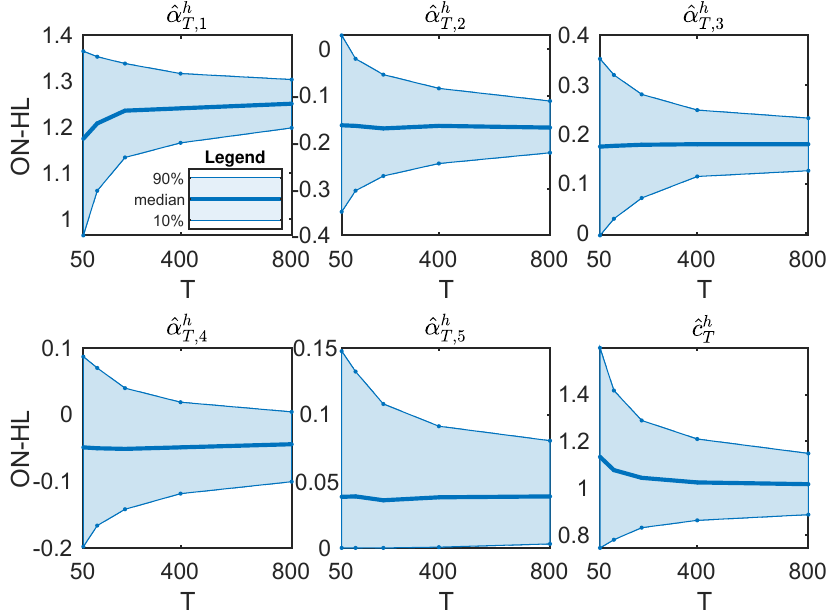}}
\end{minipage}
\caption{Quantiles of the constrained LS estimators ($\beta=5$). 
}
\label{fig:cls}
\end{figure}

{\bf Guaranteed Positivity and Stability.} 
We implement and solve the proposed SDP \clrb{for ON-HL and a constrained LS for Hawkes-Erlang} using the CVX package \cite{cvx1} in MATLAB. 
All identified HIRs are strictly non-negative and stable. 
We plot the estimator quantiles against $T$ for model order $P=5$ in Fig. \ref{fig:cls}. 
Under the conservative constraints of the Hawkes-Erlang model, the higher-order basis weights $\hat\alp_{T,j}^g$ ($j=2,3,4$) are mostly zeroed out, contributing very weakly to the overall identification. Conversely, the ON-HL framework allows for negative weights without violating non-negativity. This ensures that all basis functions remain active and effectively contribute to the LS identification. 
Higher model orders ($P \geq 10$) exhibit similar behavior but are omitted for brevity. 

\begin{figure}[t]
\hfill
\begin{minipage}[t]{1\linewidth}
\centering{\includegraphics[width=8cm]{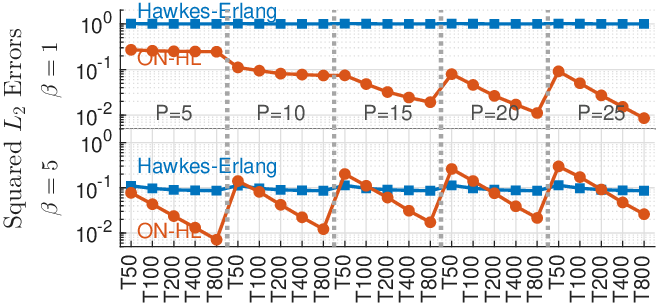}}
\end{minipage}
\caption{\clrb{The squared $L_2$ HIR approximation errors for the constrained LS.}}
\label{fig:l2err}
\end{figure}

We plot in Fig. \ref{fig:l2err} the median squared $L_2$ errors, $\int_0^{\frac{50}\beta} (\phio(t)-\phi_q(t;\hat{\alp}_T^q))^2 dt$ ($q=h$ or $g$) to quantify the HIR approximation error. 
\clrb{Here, we also considered $\beta=1$. 
Since the Laguerre basis is dense \cite{Wahl91} in $\Lw$ and $\Lt$,
 we expect that, in principle, the true HIR can be approximated arbitrarily well by increasing $P$, at the cost of requiring larger $T$. A rigorous proof is left for future work, but Fig. \ref{fig:l2err} (especially for $\beta = 1$) supports this behaviour. 
 For small orders $P = 5, 10$, the $L_2$ error saturates as $T$ grows, whereas for larger $P$ the error continues to decrease even at large $T$. 
On the contrary, the Hawkes-Erlang error is completely insensitive to model orders. 
When $T$ is fixed, high orders are not automatically preferable since 
increased model complexity requires more data to be accurately identified.
In this case, the empirically best ON-HL model orders for $\beta = 1$ are $P = 15$ at $T = 50$ and $P = 25$ at $T = 800$. For $\beta=5$, the optimal order across all $T$ is $P=5$. 

The computational time of the constrained LS estimates is independent of $T$ \cite[Section 6.4]{Bacr15}, since $\RqT,\sqT$ are precomputed. For ON-HL, the average computation times range from $0.21$ s at $P = 5$ to $0.35$ s at $P = 25$, whereas for the Hawkes-Erlang model, they remain around $0.15$ s for all $P$. 
While ON-HL requires solving an SDP, which is more complex than the affine constraints used for the Hawkes-Erlang model, its identification remains computationally efficient due to the convexity of the optimization problem.}

\mysec{Conclusions and Future Work} \label{sec:con}
We proposed a system-theoretic identification framework for Hawkes processes based on an orthonormal Laguerre expansion. We showed that the associated asymptotic Gram matrix is uniformly well-conditioned in the model order, whereas the Hawkes-Erlang Gram matrix becomes exponentially ill-conditioned. A state-space realization yields a Lyapunov equation for the empirical Gram matrix, enabling efficient LS computation for ON-HL. Using an SOS characterization of non-negative polynomials, we formulated a trace-based SDP that enforces both positivity and stability of the identified intensity. Simulations confirmed that ON-HL is numerically robust and achieves lower HIR approximation error than the Erlang-based model. Future work includes extending ON-HL to multivariate Hawkes processes and \clrb{developing its full asymptotic theory.}
\renewcommand{\thesection}{\Alph{section}}
\setcounter{section}{0}
\noleminapx
\myapp{Appendix: Proofs}
{\it Proof of Lemma \ref{lem:asym}. }
While \cite[\clrb{Theorem 7, Lemmas 13}]{Rong26a} established parallel results for a density-like basis ($q(t)\in\bR^P_{\geq0}$, $\intoi q_j(t)dt=1$), those strict conditions are unnecessary here. The proof \clrb{\cite[Theorem 7]{Rong26a}} that $R_T^q>0$ relies on the  Cauchy–Schwarz inequality, which does not depend on non-negativity or unit mass of the basis. The convergence stated in Lemma~\ref{lem:asym} follow under the weaker assumptions stated there by invoking the ergodic lemma \cite[Lemma~9]{Rong26a} and the results on vanishing pre-sample effects and martingale terms in \cite[Lemmas 10 and Lemma 11]{Rong26a}, all of which are independent of the non-negativity condition.
\hfill$\square$

{\it Proof of Theorem \ref{thm:eigR}.}
First, use the inequalities $1-|a|\leq|1-a|\leq1+|a|,a\in\bC$ and 
$|\phibo(\jw)|=|\intoi e^{-\jw t}\phio(t)dt|\leq\intoi\phio(t)dt=\Gam$ to find 
$\frac{\Lam}{(1+\Gam)^2}\leq\bar C(\ome)\leq\frac{\Lam}{(1-\Gam)^2}$. 
Since $h$ is orthonormal, we have $\intoi h(t)h(t)^\top dt = \wo{2\pi}\intoi\bar h(\jw)\bar h(-\jw)^\top d\ome=I_P$, by Parseval's theorem. 
The first set of bounds for the ON-HL model follows directly. 

For the Erlang basis, let $x_{-}$ and $y_{-}$ be the left and right 
singular vectors of $L$ corresponding to the smallest singular value 
$\sig_{\min}(L)$, respectively.
By the Rayleigh quotient, 
$\rho_{\min}(\Rsg)=\min_{\Ver{x}=1}x^\top\Rsg x = \min_{\Ver{x}=1}x^\top L\Rsh L^\top x\leq x_{-}^\top L\Rsh L^\top x_{-} = \sig_{\min}^2(L)y_{-}^\top\Rsh y_{-}\leq\sig_{\min}^2(L)\rho_{\max}(\Rsh)$, as quoted. 
The counterpart for the 
lower bound of $\rho_{\max}(\Rsg)$ follows analogously. 
Further, for the lower-triangular $L$ and $L^{-1}$,  
we have $\rho_{\max}(L)=\max_i|L_{ii}| = \sqrt{\beta/2}$ and $\rho_{\max}(L^{-1}) = \max_i |(L^{-1})_{ii}| = 2^{P-1}\sqrt{2/\beta}$. 
By Weyl's inequality, 
$\sig_{\min}(L) = \wo{\sig_{\max}(L^{-1})} \leq \wo{\rho_{\max}(L^{-1})}=2^{-(P-1)}\sqrt{\beta/2}$ and $\sig_{\max}(L) \geq \rho_{\max}(L)=\sqrt{\beta/2}$. 
Combining these bounds, the results follow.
\hfill$\square$

{\it Proof of Theorem \ref{thm:cal}. }
From Lemma \ref{lem:ss}, we have
$
	d(\chitilh(t)-\chihTh)(\chitilh(t)-\chihTh)^\top
=	A\chitilh(t)(\chitilh(t)-\chihTh)^\top dt + (\chitilh(t)-\chihTh)\chitilh(t)^\top A^\top dt 
	+ B(\chitilh(t)-\chihTh)^\top d\Ntilt + (\chitilh(t)-\chihTh)B^\top d\Ntilt
	+ BB^\top(d\Ntilt)^2. 
$
Thanks to the orderliness condition, we have $d\Ntilt\in\{0,1\}$ so that $(d\Ntilt)^2=d\Ntilt$ (See \cite{Hawk71}). 
Note that $\RhT = \intoT(\chitilh(t)-\chihTh)(\chitilh(t)-\chihTh)^\top dt=\intoT\chitilh(t)(\chitilh(t)-\chihTh)^\top dt$. Integrate both sides to get 
$(\chitilh(T)-\chihTh)(\chitilh(T)-\chihTh)^\top - \chihTh\chihThT
=AT\RhT + T\RhT A^\top + BT\shTT+T\shT B^\top + BB^\top\NtilT$. 
Divide through $T$ and reorganize to get the quoted result.
\hfill$\square$

{\it Proof of Theorem \ref{thm:sos}. }
\clrb{Part (a) follows from Lemma \ref{lem:sos}, noticing the trace identity, e.g. 
$
	\trace\{u_0^{m_1-1}(t)^\top Q_1u_0^{m_1-1}(t)\} 
=	\trace\{u_0^{m_1-1}(t)u_0^{m_1-1}(t)^\top Q_1\}
=	 \ssum{k=0}^{2m_1-1}\trace\{F_{m_1,k}Q_1\}u_k(t)
$.

For part (b), from the definition \eqref{eq:F} and the orthonormality, we first find that 
$
	\intoi u_0^{m-1}(t)u_0^{m-1}(t)^\top u_k(t) w^2(t)dt 
= 	\ssum{j=0}^{2m-2}F_{m,j}\intoi u_j(t)u_k(t)w^2(t)dt = F_{m,k}
$. 
Thus, we have $F_{m,0}=u_0(t)I_m=\sqrt{2/\beta}I_m$ and similarly 
$G_{m,k}=\intoi tu_0^{m-1}(t)u_0^{m-1}(t)^\top u_k(t)w^2(t)dt$. 
Further, use the vector three-term recursion \eqref{eq:3tr}, 
to find 
\eqn{
	G_{m,k}
=	&\mat{\intoi tu_0^{m-1}(t)u_0^{m-1}(t)^\top u_k(t)w^2(t) dt}\nonumber\\
= 	&J_m\mat{\intoi }u_0^{m-1}(t)u_0^{m-1}(t)^\top u_k(t)w^2(t)dt\nonumber\\
	&+\wo{\rho_{m-1}}\mat{\intoi} e_mu_m(t)u_k(t)w^2(t)dt\nonumber\\
=	&J_m F_{m,k}+\wo{\rho_{m-1}}\sqbra{\smat{0_{(m-1)\times m}\\F_{m+1,k}^{(m+1)}}}.\label{eq:Gmk}
} 
Now use the scalar three-term recursion in Lemma \ref{lem:3tr} to find 
\eqn{
	G_{m,k}
=	&\ssum{j=0}^{2m-1}F_{m,j}\mat{\intoi} tu_j(t)u_k(t)w^2(t)dt\nonumber\\
= 	&\wo{\rho_{k-1}}F_{m,k-1}+\frac{\kap_k}{\rho_k}F_{m,k} + \wo{\rho_{k}}F_{m,k+1}.\label{eq:Gmk2}
} 
Inserting \eqref{eq:Gmk2} into \eqref{eq:Gmk} yields the recursions.
\hfill$\square$}


\bibliographystyle{IEEEtran}
\bibliography{spHL.bib}  
\vfill\pagebreak
\end{document}